\documentclass[sigconf,balance=false]{acmart}
\usepackage{popets}

\setcopyright{popets}
\copyrightyear{YYYY}

\acmYear{YYYY}
\acmVolume{YYYY}
\acmNumber{X}
\acmDOI{XXXXXXX.XXXXXXX}
\acmISBN{}
\acmConference{Proceedings on Privacy Enhancing Technologies}
\usepackage[absolute,showboxes]{textpos} %

\usepackage[english]{babel}
\usepackage{bbding}
\usepackage{balance}
\usepackage{algorithm,algorithmic}
\usepackage{graphicx}
\usepackage{textcomp}
\usepackage{hyperref}
\hypersetup{hidelinks=true}
\usepackage{booktabs}
\usepackage{multirow}
\usepackage{scalerel}
\usepackage{tikz}
\usetikzlibrary{svg.path}
\usepackage{rotating}
\usepackage{tabularray}
\usepackage{xspace}
\usepackage{cleveref}
\usepackage{soul}
\usepackage{subcaption}
\usepackage[flushleft]{threeparttable}
\usepackage{enumitem}
\usepackage{colortbl}
\usepackage[most]{tcolorbox}

\newtcolorbox[auto counter, number within=section]{summary}[2][]{%
  colframe=blue!80!black, 
  colback=blue!10, 
  coltitle=black, 
  fonttitle=\bfseries, 
  title=Takeaway~\thetcbcounter: #2,#1,
  boxsep=1mm,   %
  left=1mm,     %
  right=1mm,    %
  top=1.3mm,      %
  bottom=1mm    %
}

\newcommand{\ie}{i.e., \@}
\newcommand{\eg}{e.g., \@}

\newcommand{\intractable}{intractable\xspace}
\newcommand{\cookiejar}{\textit{Cookie Jar}\xspace}
\newcommand{\sentcookies}{\textit{Sent Cookies}\xspace}

\newcommand{\openwpm}{OpenWPM\xspace}
\newcommand{\bannerclick}{BannerClick\xspace}

\newcommand{\regular}{\emph{Popularity}\xspace}
\newcommand{\regularreverse}{\emph{Popularity-Reverse}\xspace}
\newcommand{\random}{\emph{Random}\xspace}
\newcommand{\randomreverse}{\emph{Random-Reverse}\xspace}
\newcommand{\randcomb}{\emph{RandComb}\xspace}

\newcommand{\setter}{\textit{setter website}\xspace}
\newcommand{\sender}{\textit{sender website}\xspace}
\newcommand{\host}{\textit{tracker domain}\xspace}
\newcommand{\setters}{\textit{setter websites}\xspace}
\newcommand{\senders}{\textit{sender websites}\xspace}
\newcommand{\hosts}{\textit{tracker domains}\xspace}

\newcommand{\parax}[1]{\noindent \textbf{#1}}

\newcommand{\todo}[1]{\textcolor{blue}{#1}}

\renewcommand{\todo}[1]{#1\xspace}

\newcommand{\todoo}[1]{\textcolor{blue}{#1}}

\renewcommand{\todoo}[1]{#1\xspace}

\crefname{section}{\S}{\S}
\crefname{subsection}{\S}{\S}
\crefname{subsubsection}{\S}{\S}

\renewcommand{\shortauthors}{Rasaii et al.}

\begin{document}

\settopmatter{printfolios=true}

\setlength{\TPHorizModule}{\paperwidth}
\setlength{\TPVertModule}{\paperheight}
\TPMargin{5pt}
\begin{textblock}{0.8}(0.1,0.02)
     \noindent
     \footnotesize
	 If you cite this paper, please use the PoPETs reference:
     Ali Rasaii, Ha Dao, Anja Feldmann, Mohammadmadi Javid, Oliver Gasser, and Devashish Gosain. 2025.
     Intractable Cookie Crumbs: Unveiling the Nexus of Stateful Banner Interaction and Tracking Cookies. In \textit{Proceedings of the 25th Privacy Enhancing Technologies Symposium (PETS 2025), July 14–19, 2025, Washington, DC, USA}.
\end{textblock}

\title{Intractable Cookie Crumbs: \ Unveiling the Nexus of Stateful Banner Interaction and Tracking Cookies}

\author{Ali Rasaii}
\affiliation{%
  \institution{Max Planck Institute for Informatics}
  \state{}
  \city{}
  \country{}
  }

\author{Ha Dao}
\affiliation{%
  \institution{Max Planck Institute for Informatics}
  \state{}
  \city{}
  \country{}}

\author{Anja Feldmann}
\affiliation{%
  \institution{Max Planck Institute for Informatics}
  \state{}
  \city{}
  \country{}}
\email{}

\author{Mohammadmahdi Javid}
\affiliation{%
  \institution{Max Planck Institute for Informatics}
  \state{}
  \city{}
  \country{}
  }
\email{}

\author{Oliver Gasser}
\affiliation{%
  \institution{IPinfo}
  \state{}
  \city{}
  \country{}}
\email{}

\author{Devashish Gosain}
\affiliation{%
  \institution{IIT Bombay}
  \state{}
  \city{}
  \country{}}
\email{}

\begin{abstract}

\todo{In response to the ePrivacy Directive and the consent requirements introduced by the GDPR, websites began deploying consent banners to obtain user permission for data collection and processing.}
However, due to shared third-party services and technical loopholes, non-consensual cross-site tracking can still occur. In fact, contrary to user expectations of seemingly isolated consent, a user's decision on one website may affect tracking behavior on others.

In this study, we investigate the technical and behavioral mechanisms behind these discrepancies. Specifically, we disclose a persistent tracking mechanism exploiting web cookies. These cookies, which we refer to as \textit{\intractable}, are initially set on websites with accepted banners, persist in the browser, and are subsequently sent to trackers before the user provides explicit consent on other websites.
To meticulously analyze this covert tracking behavior, we conduct an extensive measurement study performing stateful crawls on over 20k domains from the Tranco top list, strategically accepting banners in the first half of domains and measuring \intractable cookies in the second half. Our findings reveal that around 50\% of websites send at least one \intractable cookie, with the majority set to expire after more than 10 days. 
In addition, enabling the Global Privacy Control (GPC) signal initially reduces the number of \intractable cookies by 30\% on average, with a further 32\% reduction possible on subsequent visits by rejecting the banners.
Moreover, websites with Consent Management Platform (CMP) banners, on average, send 6.9 times more \intractable cookies compared to those with native banners. Our research further reveals that even if users reject all other banners, they still receive a large number of \intractable cookies set by websites with cookie paywalls. Additionally, our measurement on the partitioned cookies---cookies that are restricted to the top-level site and thus mitigate cross-site tracking---shows that only 1.3\% of tracking cookies are marked as such, indicating their minimal impact on cross-site tracking via \intractable cookies.

 \end{abstract}


\vspace{-10mm}

\maketitle

\section{INTRODUCTION}

In the digital age, online tracking mechanisms have become a common aspect of Internet use~\cite{bielova2017web,iqbal2023tracking,munir2023cookiegraph,fouad2022my,acar2014web}. These mechanisms—such as third-party tracking cookies---form the foundation of data-driven marketing and advertising strategies. 
\todoo{Statistics show that major tech companies depend heavily on online advertising~\cite{statista_google_ad_revenue_2024, statista_meta_ad_revenue_2024}, employing various tracking techniques to monitor user activity and optimize personalized ads.}
The opaque nature of data-handling practices has led to a growing demand for transparency and control over personal data among privacy advocates~\cite{pavlou2011state, lee2011personalisation, belanger2002trustworthiness}.
In response, legislative bodies have introduced strict data protection laws, like the ePrivacy Directive~\cite{eprivacy2002directive} and the General Data Protection Regulation (GDPR)~\cite{GDPR} in the European Union and the California Consumer Privacy Act (CCPA)~\cite{CCPA} in California. 
These laws aim to limit unrestricted user data collection and enforce a more transparent consent process.

Consequently, a significant shift toward preserving user privacy can be observed. Research indicates a reduction in third-party tracking and increased visibility of privacy policies within the EU~\cite{sorensen2019before, degeling2019we, kretschmer2021cookie, dabrowski2019measuring}. 
\todo{To comply with the ePrivacy Directive’s mandate requiring prior consent for storing or accessing non-essential information on a user's device, websites increasingly deploy cookie banners to inform users and obtain their consent~\cite{degeling2019we, rasaii2023exploring}. \todoo{The GDPR further strengthened and enforced the standards and legal criteria for valid consent established under the ePrivacy Directive.} 
Despite these advancements, previous studies~\cite{trevisan2019, Matte2020respect, rasaii2023exploring} demonstrate that websites continue to set tracking cookies extensively, even without explicit user consent. While the mere act of placing such cookies is not necessarily a violation of privacy regulations—and, in some cases, may be justified under Article 6 GDPR—the fact that most of these cookies originate from well-known tracking companies raises concerns about the alignment of such practices with the principles of transparency, purpose limitation, and user autonomy.}

Furthermore, the design and implementation of consent banners have been criticized for employing manipulative tactics that skew user behavior, often leading to uninformed or coerced consent decisions~\cite{kosta2013peeking, machuletz2020multiple, soe2020circumvention}. 
For instance, users may accept cookie banners when there is no easy refusal option, resulting in the setting of tracking cookies.
These unintentionally accepted cookies may increase the likelihood of being tracked during future web browsing. 
\todo{The stateful cross-site reuse of tracking cookies before obtaining user consent is a previously under-explored and unmeasured threat. Exploring cookie banners and their impact in a stateful manner is therefore crucial, as both browsing and tracking unfold over time, often involving interconnections across multiple sites and entities.}

In this paper, we uncover a persistent inter-domain tracking mechanism via web cookies. Specifically, we find that numerous tracking cookies, \textit{initially set on a website with an accepted banner, continue to be transmitted to tracker domains even before users interact with the rejectable banners} on other websites.
Unlike previous studies that focused on the \textit{stateless deployment} of cookies on websites, we conduct a \textit{stateful} banner interaction across websites, demonstrating how seemingly opposing decisions on one website (with accepted banner) can influence tracking behavior on subsequent websites (with rejected banner), resulting in the \textit{transmission} of cookies to trackers.
This not only violates privacy regulations such as the GDPR by undermining the overall effectiveness of banners as the primary consent mechanism, but also creates a false sense of privacy when users reject the banners.
Throughout this paper, we refer to these cookies as \textbf{\intractable cookies} (see \Cref{sec:intractablecookie}).

To substantiate our findings, we conduct two measurement campaigns using \bannerclick~\cite{rasaii2023exploring}, a tool designed to automatically detect and interact with cookie banners. We improved its performance, particularly by increasing its rejection accuracy from 87\% to 99\% (see \Cref{sec:meth}). In the first campaign, we crawl the top 20,000 websites from Tranco, accepting cookie banners in the first half and measuring the number of \intractable cookies on successfully rejected domains in the second half.
For the second campaign, we randomly select and shuffle 20k sites from Tranco's top 50k, following a similar approach. Additionally, we perform both runs in reverse order to gain further insights.
Overall, our main findings can be summarized as follows:
\begin{itemize}[leftmargin=*]
    \item We find that nearly 50\% of websites send at least one \intractable cookie to third-party tracking domains before obtaining explicit user consent (see \Cref{subsec:cookie-distribution}).
    \item Regarding the effect of banner interaction, we see no immediate changes on \intractable cookies after rejecting banners. However, on average, 25\% of \intractable cookies are not sent after reloading the webpage with the rejected banner (see \Cref{subsec:intract}).
    \item Furthermore, our measurements show that enabling the Global Privacy Control (GPC) signal in the browser can initially reduce \intractable cookies by an average of 30\%. An additional 32\% reduction is achievable on subsequent visits by also rejecting cookie banners (see \Cref{subsec:gpc}).
    \item We observe that more popular Tranco websites send fewer \intractable cookies compared to less popular ones. Specifically, the top 50 websites send, on average, zero \intractable cookies, while the top 10k websites send 25 \intractable cookies (see \Cref{subsec:ranking}).
    \item We note that websites using CMP banners send more than 6.9 times as many \intractable cookies compared to those using native banners. Moreover, out of 5,915 accepted domains, 90 websites with cookie paywalls 
    are responsible for setting more than 35\% of \intractable cookies (see \Cref{subsec:banner}).
    \item We also observe $\approx60\%$ of \intractable cookies have an expiration time of at least 10 days, highlighting their persistence. In addition, around 90\% of \intractable cookies are set (or refreshed) by at most 1\% of accepted websites (see \Cref{subsec:cookie-exp-dup}).
    \item Our analysis shows that, on average, each domain has 3.42 different trackers, with each tracker receiving an average of 7.3 \intractable cookies. We also verified that the top 20 trackers are indeed well-known tracking companies (see \Cref{subsec:domain-analysis}).
    \item Additionally, to assess the impact of partitioned cookies\todo{---whose transmission is restricted to their \setter---}in mitigating \intractable cookies, we conduct a separate measurement using Chrome. The results show that only 1.3\% of all unique tracking cookies are partitioned, with more than half accompanied by non-partitioned cookies from the same tracker domain (see \Cref{subsec:partitioned}).

\end{itemize}

Finally, our findings reveal a persistent gap between privacy regulations, such as the GDPR and ePrivacy Directive, and their technical implementation. In \Cref{sec:discuss}, we examine how fragmented interpretations and unclear accountability contribute to this gap, and introduce our browser-integrated approach as a potential solution.

\section{BACKGROUND AND RELATED WORK}

\definecolor{darkgreen}{rgb}{0.0, 0.5, 0.0}  %
\definecolor{darkred}{rgb}{0.7, 0.0, 0.0}    %

\begin{table}[t]
    \centering
    \small
    \resizebox{\linewidth}{!}{
    \begin{tabular}{lrrrr}
    \toprule
    \textbf{References} & \textbf{Automated} & \textbf{Reject Coverage} & \textbf{Stateful} & \textbf{Sent Cookie} \\
    \midrule
    Englehardt et al.~\cite{englehardt2016online} & \textcolor{darkgreen}{\Checkmark} & N/A & \textcolor{darkgreen}{\Checkmark} & \textcolor{darkred}{\XSolidBrush} \\
    Trevisan et al.~\cite{trevisan2019} & \textcolor{darkgreen}{\Checkmark} & CMP & \textcolor{darkred}{\XSolidBrush} & \textcolor{darkred}{\XSolidBrush} \\
    Matte et al.~\cite{Matte2020respect} & \textcolor{darkred}{\XSolidBrush} & CMP & \textcolor{darkred}{\XSolidBrush} & \textcolor{darkred}{\XSolidBrush} \\
    Jha et al.~\cite{jha2022internet} & \textcolor{darkgreen}{\Checkmark} & N/A & \textcolor{darkred}{\XSolidBrush} & \textcolor{darkred}{\XSolidBrush} \\
    Smith et al.~\cite{Smith2024GDPRCompliance} & \textcolor{darkgreen}{\Checkmark} & CMP & \textcolor{darkred}{\XSolidBrush} & \textcolor{darkred}{\XSolidBrush} \\
    Rasaii et al.~\cite{rasaii2023exploring} & \textcolor{darkgreen}{\Checkmark} & 87\% & \textcolor{darkred}{\XSolidBrush} & \textcolor{darkred}{\XSolidBrush} \\
    \midrule
    \textbf{Our work} & \textcolor{darkgreen}{\Checkmark} & 99\% & \textcolor{darkgreen}{\Checkmark} & \textcolor{darkgreen}{\Checkmark} \\
    \bottomrule
    \end{tabular}
    }
\caption{Overview of the closest previous studies on the misbehavior of tracking cookies. Our study is conducted in a fully automated manner, capable of rejecting 99\% of all banners. Furthermore, it is the first study to statefully measure the act of sending cookies to the tracker rather than their setting in the browsers.}
    \label{tab:related_works}

    \vspace{-5mm}
\end{table}

Statistics~\cite{statista_google_ad_revenue_2024, statista_meta_ad_revenue_2024} show that major tech companies rely heavily on online advertising, with Google generating 77\% and Facebook 98.4\% of their revenue from ads. These companies employ various tracking techniques, \eg web cookies, to collect users' online activity and deliver targeted ads while optimizing ad recommendations.
Subsequently, previous studies~\cite{pavlou2011state,  lee2011personalisation, belanger2002trustworthiness} show that people are increasingly concerned about how their personal information is being collected and used by companies.
In recent years, several data protection laws have been enacted to regulate the use of web cookies and other tracking and profiling techniques, such as the ePrivacy Directive~\cite{eprivacy2002directive} and the General Data Protection Regulation (GDPR)~\cite{GDPR} in the European Union and the California Consumer Privacy Act (CCPA)~\cite{CCPA} in California. 

\todo{The GDPR requires websites to obtain user consent before collecting or processing personal data, unless the processing is justified by another legal basis—such as being strictly necessary for the performance of a service requested by the user (Article 6).
As a result, websites increasingly rely on cookie banners to inform users about data practices and to obtain consent. Many outsource this functionality to Consent Management Platforms (CMPs)—third-party services that provide ready-to-use, configurable consent interfaces. A widely adopted framework for implementing these banners is the IAB Europe Transparency and Consent Framework (TCF)~\cite{tcf}, which standardizes how CMPs operate and transmit consent signals. While the TCF is presented as a GDPR-compliant solution to facilitate user consent and regulatory compliance, it originates from IAB Europe—an industry group representing the online advertising sector. As such, CMPs built on the TCF may prioritize maximizing consent rates, aligning with the interests of advertisers rather than promoting user privacy.}

Numerous studies have examined compliance issues, particularly identifying potential legal violations in cookie banner implementation and consent storage. While these studies show that the GDPR has led to a per-site reduction in third-party tracking and improved the visibility of privacy policies and cookie banners across the European Union, violations persist~\cite{sorensen2019before, degeling2019we, kretschmer2021cookie, dabrowski2019measuring}.
In addition, some studies have evaluated the influence of consent banner design on user behavior, specifically their acceptance or denial of consent~\cite{kosta2013peeking, machuletz2020multiple,soe2020circumvention,bielova2024effect}. 
Santos et al.~\cite{Santos2021Banners} analyzed the clarity of cookie banners and found that 61\% of them employed vague language, failing to specify privacy practices adequately. 
Utz et al.~\cite{utz2019informed} explored additional factors influencing user consent, such as banner placement, and reported significant impacts on consent decisions based on these elements. Additionally, Nouwens et al.~\cite{10.1145/3313831.3376321} showed that merely removing the opt-out button from the first layer of banners increases consent rates by approximately 23\%.
Overall, these studies consistently identified interface interference as a key factor that significantly influences how users interact with banners.

Finally, many studies specifically analyzed the techniques and prevalence of tracking cookie usage in web tracking.
\Cref{tab:related_works} compares some of the most relevant studies~\cite{trevisan2019, Matte2020respect, jha2022internet, rasaii2023exploring, Smith2024GDPRCompliance, englehardt2016online} with our work based on key distinguishing factors, such as whether they were conducted manually or through automation, their rejection coverage, whether they employ stateful or stateless measurements, and whether they focus on the setting of cookies on browsers or sending them to tracker domains. 

Englehardt et al.~\cite{englehardt2016online} conducted one of the first fully automated, large-scale studies on tracking cookies. They introduced \textit{OpenWPM}, a measurement platform capable of collecting HTTP requests and responses, JavaScript calls, and script files. Their analysis included an ID cookie detection method to uncover cookie syncing across sites. \todoo{However, even though they employed a stateful approach, their analysis relied on identifying user IDs previously set in cookies and later used in referer headers and request URLs; they did not examine the \texttt{Cookie} HTTP header to observe cookies being sent directly to servers. Nevertheless, the study was conducted in January 2016, prior to the adoption of the GDPR, and it did not explicitly discuss the implications of its findings in the context of existing regulations at the time, such as the ePrivacy Directive, nor did it consider the effect of banner interaction.}
Trevisan et al.~\cite{trevisan2019} developed CookieCheck, a tool that visits websites as a new user and analyzes cookies placed in the browser.
It focused solely on profiling cookies set by CMPs that violate the ePrivacy Directive before any user consent is given.
\todo{Matte et al.~\cite{Matte2020respect} conducted semi-automatic crawl campaigns to detect suspected GDPR and ePrivacy Directive violations in banners based on the Transparency and Consent Framework developed by IAB Europe.}
Jha et al.~\cite{jha2022internet} attempted to interact with cookie banners in an automated manner to observe differences in the cookies set. \todo{However, their work focused solely on banner acceptance in a stateless manner.} Smith et al.~\cite{Smith2024GDPRCompliance} specifically investigated the placement of tracking cookies under the guise of legitimate interest by CMPs, as well as their compliance with properly transmitting users' choices through TCF consent strings.
Finally, Rasaii et al.~\cite{rasaii2023exploring} developed BannerClick, a tool capable of both accepting and rejecting consent banners.
They conducted a comprehensive measurement study on Tranco's Top 10K sites, analyzing cookies deployment in a stateless manner.

\todo{In this research, we investigate how cookies set on a website with an accepted banner may contribute to user tracking on subsequent visited websites before any consent is given.
Among the available tools capable of interacting with banners~\cite{nouwens2022consentomatic, bouhoula2024automated}, we use an improved version of BannerClick due to its higher rejection coverage and integration with OpenWPM. This integration enables us to perform a combination of stateful and stateless crawls, capturing all the data required. We then examine the implications of our findings concerning privacy regulations and potential violations. To the best of our knowledge, this specific form of user tracking via web cookies has not been previously explored in the literature.}

\section{INTRACTABLE COOKIES}
\label{sec:intractablecookie}

\begin{figure}[t]
\centering
\includegraphics[width=0.96\columnwidth]{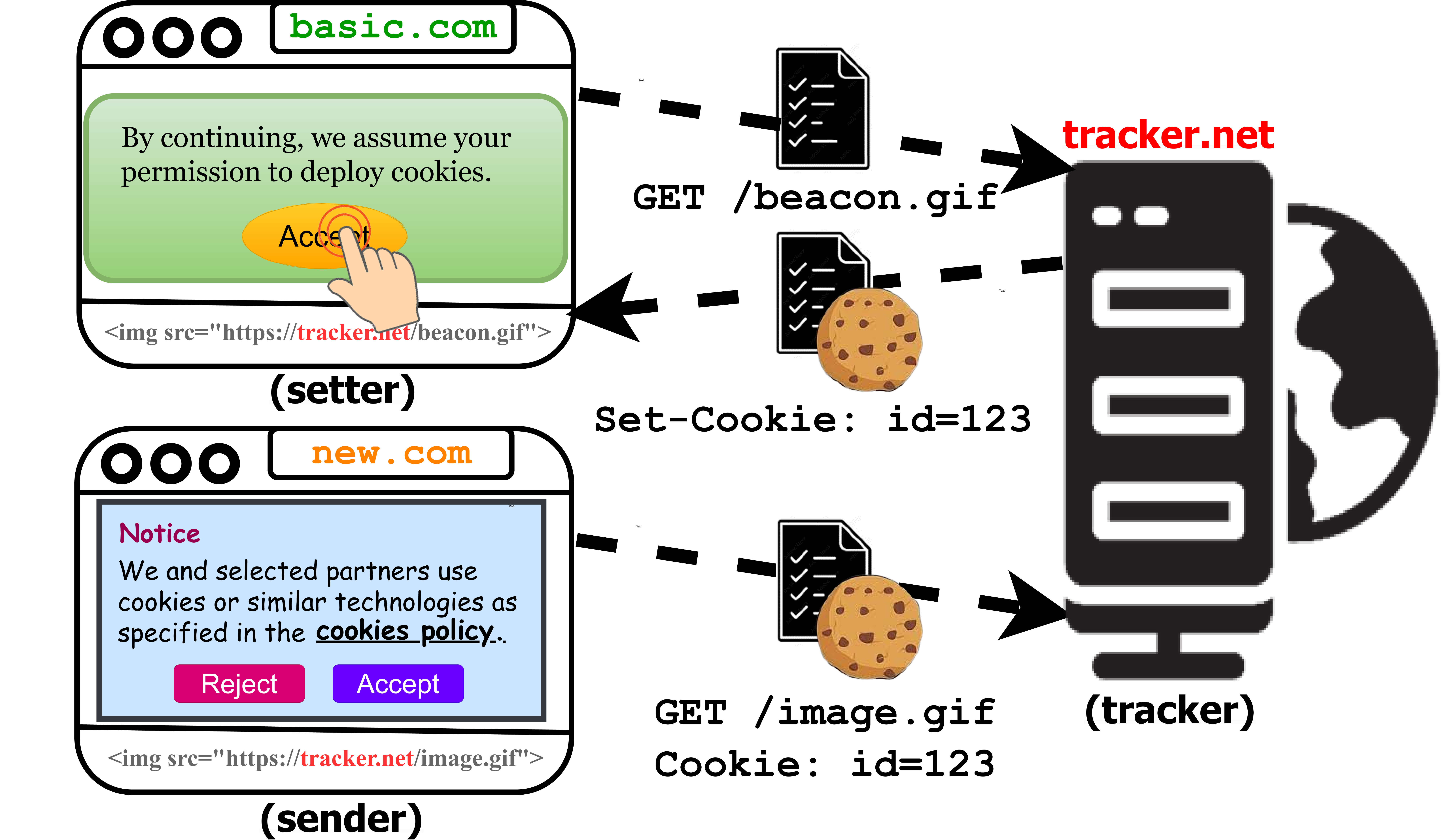}
\caption{Overview of a scenario illustrating the interconnections among the four main entities involved in \intractable cookie transmission: \setter with an accepted banner, \host, \sender with a rejectable cookie banner, and the browser.}
\label{fig:schematic}
\end{figure}

In this section, we detail the nature of \intractable cookies and the entities involved, as illustrated in \Cref{fig:schematic}. 

In the depicted scenario, a user initially visits \texttt{basic.com} and ``accepts'' the cookie banner. This might result in setting new cookies in the browser. For example, the browser starts sending HTTP requests to load a third-party resource, \texttt{beacon.gif}, from \texttt{tracker.net}, which sets a cookie ($id{=}123$). We refer to \texttt{basic.com} as the \setter, as it establishes the context for the initial cookie-setting in the user's browser, and to \texttt{tracker.net} as the \host, as it tracks the user.
Later, the user accesses \texttt{new.com}, which shows a banner with the ``reject'' option.\footnote{We focus on websites with rejectable banners when measuring \intractable cookies on \senders. This is because banners—and the associated tracking behavior—are directly shaped by privacy regulations such as the GDPR. A core requirement of these regulations is that users must be given the ability to reject cookies. Consequently, websites whose banners lack a reject option are excluded from our study.}
In this scenario, even though the user has not yet interacted with the banner to explicitly consent to the cookies, during the rendering of the webpage, the browser might send an HTTP request to load third-party resources. For example, \texttt{new.com} might embed a resource (\texttt{image.gif}) from \texttt{tracker.net}, resulting in sending an HTTP request along with the previously stored cookie ($id{=}123$) back to the \host.
In this context, \texttt{new.com} is referred to as the \sender, as it leads to sending the cookie to the \host. \todo{Ultimately, we define \textbf{\intractable cookies} as those that are initially set on a website where the user accepts the banner (\ie, the \setter), and are later sent by a \sender to the \host before any banner interaction.
}

It should be noted that existing research primarily assesses the setting of tracking cookies prior to explicit consent in a stateless manner. However, in our study, we introduce \intractable cookies to address two key factors overlooked by previous work. First, our measurements are conducted in a stateful manner—an essential perspective, as internet browsing and tracking inherently occur across sessions and time, with websites often interacting with one another. Second, we focus on the transmission of cookies rather than solely their deployment in browsers. \todo{While the GDPR governs the processing of personal data and requires a lawful basis such as consent, the act of setting a cookie may not, in itself, violate the GDPR unless personal data is involved. Though under the ePrivacy Directive (Article 5(3)), storing or accessing tracking cookies on a user’s device without prior consent constitutes a violation, regardless of whether the data qualifies as personal.}
In addition, as demonstrated in \Cref{app:cookie-jar}, the setting of \intractable cookies by \senders\footnote{For instance, visiting \texttt{new.com} might also set the cookie ($id{=}123$), which could be considered more detrimental than merely sending it.} is relatively uncommon. \todoo{In many cases, \intractable cookies are transmitted to third-party trackers without being explicitly set again.}
Therefore, treating a website as a single, isolated entity and analyzing the deployment of tracking cookies in a stateless manner may fail to capture the full extent of unwanted tracking practices in the wild and subsequently hinder regulators from crafting the most effective regulatory measures.

In \Cref{sec:discuss}, we elaborate more on the relation between the intractable cookies and privacy regulations, as well as possible mitigation approaches like partitioned cookies \cite{CookiesH71:online} and their challenges.

\section{METHODS}
\label{sec:meth}

\todo{We now describe our data collection approach, 
our crawling setup and tools, and the cookie classification method, followed by a discussion of measurement limitations.}

\begin{figure*}[t!]
\centering
\includegraphics[width=0.70\textwidth]{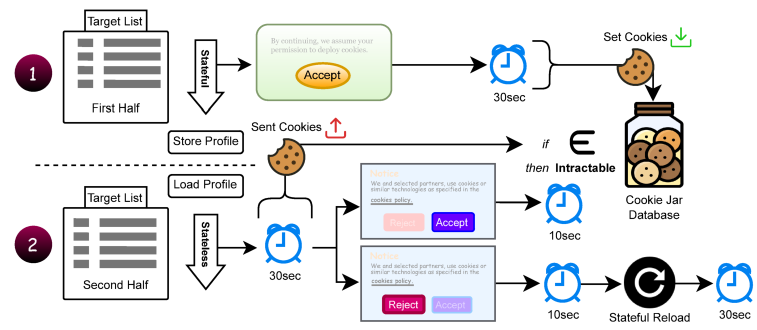}
\caption{Overview of our methodology: Each run consists of two phases. In the first, banners for half of the target list are accepted in a stateful manner. In the second phase, the browser profile from the first phase is loaded, and two iterations are performed on the remaining domains: one accepting and one rejecting the banners.
Cookies set during the first phase on accepted domains and sent to trackers in the second phase before rejecting banners are classified as \emph{intractable cookies}.}
\label{fig:method}
\end{figure*}

\subsection{Data Collection}
\label{meth:collection}

\todo{To overcome the shortcomings of existing popular site lists—such as instability, unreachable domains, and susceptibility to manipulation by adversaries~\cite{Tranco}—we utilize the Tranco website popularity ranking.\footnote{\todo{This list, generated on 07 December 2023 with ID K2NZW, is available at: \url{https://tranco-list.eu/list/K2NZW}.}}
Our target list consists of the landing pages of top-ranked domains.}
To conduct large-scale automated crawls and collect data for analysis, we use a modified version of \bannerclick~\cite{rasaii2023exploring}, a tool that effectively detects and interacts with cookie banners, \todo{and specifically identifies CMP and cookie paywall banners. \bannerclick is built on top of OpenWPM~\cite{englehardt2016online} (version 0.26.0), which uses Firefox v121.0 with TCP disabled~\cite{mozilla_tcp_2021}. OpenWPM is widely used in privacy measurement studies and enables the collection of cookies, JavaScript function calls, web resources, and HTTP/HTTPS requests and responses for each site.}

\Cref{fig:method} depicts the overview of our methodology setup. Each measurement consists of two phases: stateful and stateless. In the stateful phase, BannerClick crawls the first half of the target list. Upon successful \textit{acceptance} of a banner, we aggregate the corresponding cookies, along with their \setter, into a database called \cookiejar. At the end of this phase, we store the browser profile to use as the base profile in the next phase.

\begin{table}[t!]
    \centering
    \small
    \resizebox{\linewidth}{!}{
    \begin{tabular}{lrrrrr}
    \toprule
\textbf{Run} & \textbf{\#Crawls} & \textbf{\#Acc} & \textbf{\#Rej} & \textbf{Date} & \textbf{Duration} \\
        \midrule
   Popularity & 20k & 3,034 & 2,379 & Jul 2024 & 14 days \\
   Popularity-Reverse & 20k & 3,060 & 2,424 & Jul 2024 & 14 days \\
        \hline
   Random & 20k & 2,933 & 2,578 & Jul 2024 & 15 days \\
   Random-Reverse & 20k & 2,983 & 2,543 & Jul 2024 & 15 days \\
   RandComb & \multicolumn{1}{c}{\hspace{2.1em}--}

 & 5,916 & 5,121 & \multicolumn{1}{c}{\hspace{0.7em}--} & \multicolumn{1}{c}{\hspace{1.1em}--} \\
        \hline
   GPC-Enabled & 5,121 & \multicolumn{1}{c}{--} & 4,947 & Aug 2024 & 5 days \\
   Partitioned Cookies & 6,518 & 4,124 & \multicolumn{1}{c}{--} & Jan 2025 & 11 days \\
    \bottomrule
    \end{tabular}
    }   
    \caption{\todo{Overview of different measurement types. The columns respectively show the total number of unique domains that are crawled, accepted, and rejected, as well as the date and duration of each run. Note that the \randcomb run is a combination of the \random and \randomreverse runs.}}
    \label{tab:measurement_campaigns}
\end{table}

Next, \bannerclick crawls the second half of the target list in a stateless manner. For each crawl, it loads the final browser profile from the first phase, making the crawl stateful with respect to the accepted domains.\footnote{\todo{Throughout this paper, ``accepted domains'' refer to domains whose banners have been successfully accepted; the same applies to ``rejected domains''.}}
After accessing the domain, it waits for 30 seconds and collects all cookies sent through HTTP requests. These cookies are referred to as \sentcookies and are later compared with those in the \cookiejar to detect \intractable cookies.
In this step, BannerClick interacts with the banner in two separate iterations: one for rejecting and one for accepting. Following each interaction, it waits 10 seconds to capture any immediate changes. Then, to assess if rejection reduces the number of \intractable cookies on subsequent visits, it reloads the webpage and waits another 30 seconds. The reload event is conducted in a stateful manner, with browser caching disabled to ensure any inconsistencies are related to banner rejection.

\todo{We conduct two main measurement campaigns to analyze \intractable cookies, along with two additional runs to investigate the effects of the GPC signal and partitioned cookies. All runs are executed from a server located in the EU, and each crawl within a run is performed once without repetition.
\autoref{tab:measurement_campaigns} provides an overview of these runs:}
    
\begin{enumerate}[leftmargin=*] 
    \item \textbf{Popularity runs}:
In this measurement campaign, we use Tran\-co’s top 20k websites, first attempting to ``accept'' banners from the top 10k websites and then detecting \intractable cookies on the bottom 10k. This process is conducted with swapped domain lists to explore the impact of website rankings. We refer to these runs as \regular and \regularreverse.
 
\item \textbf{Random runs}: To further assess the \intractable cookies on a randomized selection of domains, we sample and mix up 20k domains from Tranco's top 50k domains list. This shuffled list is then split and examined in two sets, as in the popularity runs. We call these runs \random and \randomreverse.
\todo{Moreover, as shown in \Cref{subsec:cookie-distribution}, both random runs exhibit similar behavior. Thus, for a more comprehensive analysis of the impact of banner interaction, website rank, and banner types on \intractable cookies, we use a combined version of them, collectively referred to as the \randcomb run.}

\item \todo{\textbf{GPC-Enabled run}: We conduct separate stateless runs to measure the impact of the Global Privacy Control (GPC) signal on preventing \intractable cookies. Using browser profiles saved during the stateful phase of the \random and \randomreverse runs as the base, \bannerclick enables the GPC signal and attempts to reject banners previously identified as rejectable in the random runs (\ie 5,121 in total). We then combine the results of both runs—similar to the \randcomb run—and use them for the GPC analysis presented in \Cref{subsec:gpc}.}

\item \todo{\textbf{Partitioned Cookies run}: In addition, to assess the prevalence and effectiveness of partitioned cookies in mitigating \intractable cookies by limiting their transmission to the \setters, we conduct a separate stateful run using Chrome on the domains with previously detected banners on them in the \texttt{\randcomb} run (\ie 6,518 in total), accepting their banners and collecting all cookies set during the visits.
Since OpenWPM only supports Firefox and, unlike Chrome, Firefox does not expose the \texttt{partitioned\_key} field in its cookie storage, we modified \texttt{\bannerclick} to operate in Chrome’s default mode. This setup allows us to collect and measure partitioned cookies independently of OpenWPM.
The results of this measurement, along with additional details on the role of partitioned cookies in mitigating cross-site tracking, are presented in \Cref{subsec:partitioned}.}

\end{enumerate}

\todo{Throughout our analysis, to eliminate the potential impact of differing numbers of accepted domains on the cookies stored in the \cookiejar, we randomly sample and normalize the number of accepted domains when comparing two runs. Specifically, for the run with more accepted domains, we construct a new version of the \cookiejar that includes only cookies whose \textit{setters} belong to a randomly selected subset of accepted domains. The size of this subset matches the number of accepted domains in the run with fewer acceptances. For example, when comparing the \regular and \regularreverse runs, we sample 3,034 out of 3,060 accepted domains in the \regularreverse run and retain only those cookies in its \cookiejar whose \setter is among the sampled domains. This ensures that both \cookiejar instances are derived from the same number of accepted domains, thereby mitigating potential bias when comparing sent cookies.}

\noindent\todo{\textbf{Crawl Coverage:} Out of 20,000 crawls of unique domains in the \random run, 16,296 pages were successfully loaded, 692 triggered timeouts, 219 threw exceptions (e.g., due to errors during banner detection or interaction), and 2,783 were completely unreachable.} \todoo{The \randomreverse run exhibits similar behavior.}

\subsection{Modifications to OpenWPM}
We use OpenWPM~\cite{englehardt2016online} to conduct a combination of stateful and stateless crawls aiming to observe the interplay of the \textit{setter} and the \sender as mentioned earlier. OpenWPM triggers a new event whenever cookies are set, altered, or deleted and stores the corresponding data in the database.
We adapted its functionality to ensure that every event involving the addition or updating of a cookie results in overwriting the existing cookie in the browser with a new fixed expiration time (Saturday, 01 Jan 2028, 12:12:12 GMT). This modification is crucial because many cookies might otherwise expire before the completion of our measurements (see ~\Cref{subsec:cookie-exp-dup}).
Moreover, a cookie can be deleted by overwriting its expiry time before the current time. We identify such cases and allow them to proceed without extending the expiration time. 
In addition, we frequently store the browser profile to resume from the last saved profile upon unexpected crashes. 

Furthermore, we integrated a request parser within OpenWPM that processes HTTP requests by parsing the headers and storing all associated cookies along with the corresponding \sender. This enhancement facilitates the comparison of cookies retrieved from HTTP requests on domains where banners are rejected (\ie \sentcookies) with those stored in \cookiejar. 

\subsection{Modifications to BannerClick} 
\label{meth:bannerclick}

To cover a broader range of banners in our study, we enhanced BannerClick's accuracy in detecting and interacting with banners, particularly newer types of banners that it was previously unable to handle. \todo{In particular, we observed that many banners now provide buttons allowing users to accept (or save) the granular options through the ``Settings'' layer. Since these granular options typically preselect only essential cookies, clicking these buttons is expected to reject tracking cookies. See \Cref{app:banners} for an example of such banners.}
To handle these cases, we modify BannerClick so that if it cannot find the ``Reject'' button in the banner's main layer, it attempts to click the ``Settings'' button. 
\todo{If successful, it then attempts to click ``Reject All'' buttons (or semantically similar ones such as ``Decline All,'' including equivalents in other languages) within the ``Settings'' layer. If the ``Settings'' layer remains visible in the viewport, \bannerclick then tries to detect and click buttons containing text such as ``Confirm,'' ``Save,'' ``Accept Selected,'' or their semantic or linguistic equivalents.}

\begin{figure*}[t!]
     \centering
     \begin{subfigure}[b]{0.490\linewidth}
         \centering
         \includegraphics[width=\textwidth]{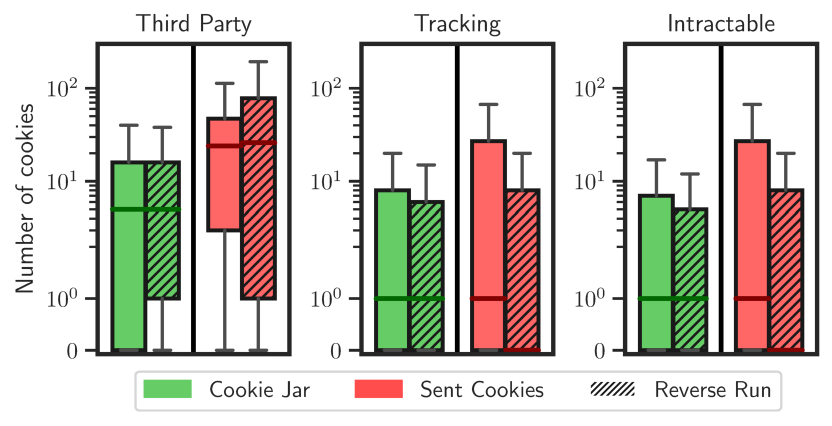}
         \caption{Popularity and Popularity-Reverse runs.}
         \label{fig:box_regular}
     \end{subfigure}
     \hfill
          \begin{subfigure}[b]{0.490\linewidth}
         \centering
         \includegraphics[width=\textwidth]{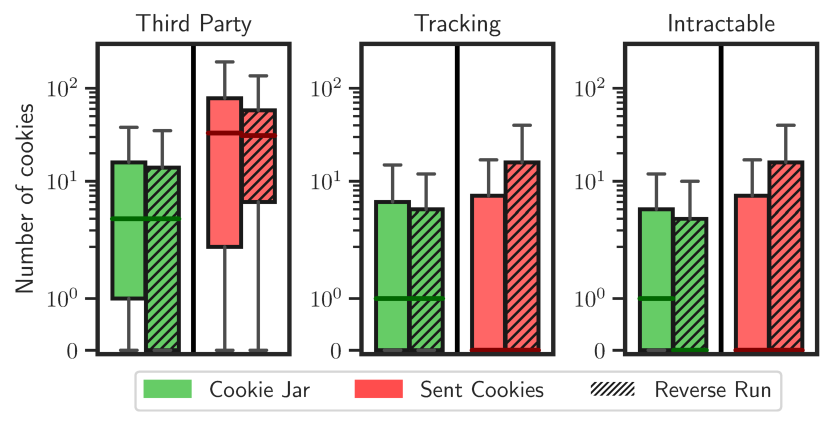}
         \caption{Random and Random-Reverse runs.}
         \label{fig:box_random}
     \end{subfigure}
        \caption{Cookie distribution for popularity and random runs.}
        \label{fig:cookie_distribution}
\end{figure*}

To evaluate the accuracy of banner rejection after the changes, we randomly selected 1k domains from the top 20k in the Tranco list and attempted to reject their banners. Out of 351 websites with detected banners, BannerClick successfully rejects 263 of them, of which 39 are rejected by clicking on the ``Settings'' option (20 with preselected options and 19 with a ``Reject All'' option), while the remaining 224 are directly rejected.
We manually reviewed the screenshots and found that only one is a false positive and four are false negatives,\footnote{The false positive occurs when clicking a ``Não, ajustar'' button that opens the ``Settings'' layer (\ie \texttt{kinghost.com.br}). False negatives are mainly due to language limitations and uncommon cookie banner designs, such as slide-out panels triggered by the ``Settings'' button (\eg \texttt{edg.io}).} resulting in improving the rejection accuracy from 87\%~\cite{rasaii2023exploring} to around 99\%. Note that among the rejected banners of our sample set, there are no cases where the preselected preferences include non-essential cookies. We publish the modified versions of OpenWPM and BannerClick, as well as analysis scripts and data at \texttt{bannerclick.github.io}.

\subsection{Cookie Classification}
\label{subsec:classify}
To categorize cookies as first-party or third-party, we utilize the latest Public Suffix List~\cite{public-suffix-list} to determine the effective top-level domains (eTLD+1) of both the visited websites and the cookies' \texttt{host} attribute. We then compare the domain of each cookie to the domain of its \setter. If they match, the cookie is classified as first-party; if not, it is deemed third-party.

Furthermore, we utilize the \emph{justdomains} blocklist~\cite{justdomains} to identify \hosts. This list aggregates domain entries from several widely used filter lists, including EasyList, EasyPrivacy, AdGuard, and NoCoin, and has been adopted in prior studies~\cite{gotze2022measuring, rasaii2023exploring, rasaii2023thou}.
To obtain the most recent version (\ie February 2025), we convert the blocklists into the equivalent justdomain lists using the JustDomain converter script~\cite{justdomains_ci}. 
Next, we compare the domains in the lists with the host attribute of the cookies. If there is an exact match or if the host ends with a domain from the list, preceded by a period (`.'), we classify the cookie as a tracking cookie.\footnote{We iterate over all entries in the filter lists and compare them with the host of cookies using the following condition:\\ \texttt{if host == entry or host.endswith('.' + entry) then True}.}

Subsequently, we categorize a tracking cookie as \emph{\intractable} if it is set on a website with an accepted banner during the first phase of the run (\ie the stateful phase), and later sent to the tracker from a website in the second phase (\ie the stateless phase) before its banner is successfully rejected.

\subsection{Measurement Limitations} 
Despite our best efforts to eliminate bias from our measurements, we acknowledge that our study does not fully capture the variety of real-world scenarios users encounter while browsing. 
First, the individual browsing behaviors are far more complex than the direct crawling of a list of domains. Second, website responses may vary based on several factors, including user activities such as scrolling, and variations in browser settings and capabilities. For example, a study shows that further user interaction and deeper crawling can lead to a 36\% rise in the use of tracking technologies like cookies \cite{Urban_2020}. \todo{Prior work also shows that websites can exhibit varying behavior across crawls, even when performed within a short time frame. Moreover, factors such as the user agent and whether the crawler lands on the homepage or an inner page can affect the results~\cite{rasaii2023exploring, 10.1145/3618257.3624795}. Additionally, our artificial extension of cookie expiration times may introduce some bias; however, as discussed in \Cref{subsec:cookie-exp-dup}, this effect is minimal.}

Furthermore, classifying cookies as trackers presents inherent challenges due to the lack of definitive references on their actual usage. Among the many available approaches, each with its own limitations and advantages~\cite{munir2023cookiegraph, gundelach2023cookiescanner, chen2021cookie}, we identify tracking domains using the justdomains block filter list due to its widespread use and comparability with related studies. 
However, these lists are crowdsourced, meaning they are continuously updated and maintained by volunteers. As a result, they may overlook certain cases or contain misconfigurations in their exception procedures.  
For instance, during manual inspection, we identified several third-party cookies classified as \intractable cookies, owned by well-known trackers such as \texttt{doubleclick.net} and \texttt{demdex.net}, that were not included in the final justdomains lists.  
In particular, we found that \texttt{doubleclick.net} used a cookie named \texttt{IDE} as an \intractable cookie on 2,300 rejected websites. According to Google documentation~\cite{GoogleAdsCookies}, this cookie is used to record users' interactions with websites' front-end to enable personalized advertising.
\todoo{Moreover, we use a domain-only filter list instead of a full rule-based filter list because our focus is on identifying data transmissions to known \hosts, rather than directly measuring or attributing tracking cookies. While this simplified approach may miss some tracking cookies, it still allows us to capture the broader network interactions with tracker infrastructure.}

\section{RESULTS}
\label{sec:results}

\todo{In this section, we present our results and analyze their implications. First, we measure the prevalence of \intractable cookies and examine how they are affected by banner interactions and the GPC signal. We then assess how factors such as website ranking and banner type influence the deployment and transmission of \intractable cookies. Finally, we investigate their persistence, association with major trackers, and the potential effectiveness of partitioned cookies in mitigating third-party cross-site tracking via \intractable cookies.}

\subsection{Cookie Distribution}
\label{subsec:cookie-distribution}

\Cref{fig:cookie_distribution} shows the overall distribution of third-party, tracking, and \intractable cookies across popularity and random runs. Cookies set in accepted domains in the stateful phase (\ie \cookiejar) and those sent via HTTP headers before the successful rejection of banners in the stateless phase (\ie \sentcookies) are depicted using green and red boxplots, respectively. Note that, for measuring \intractable cookies in \cookiejar (\ie \intractable category green boxplots), we check if a cookie with the same \texttt{name} and \texttt{host} (the \setter might be different) is later sent as \intractable cookies. Therefore, when we refer to a cookie in \cookiejar as \intractable, it does not imply that it is inherently an \intractable cookie. In other words, \intractable cookies are essentially third-party tracking cookies that are set with user permission on one website but later sent and propagated covertly without user permission to other websites across stateful browsing sessions, ultimately subverting the core purpose of cookie banners.

In \Cref{fig:box_regular}, for tracking cookies, we see nearly identical \cookiejar boxplots (with median one) across both \regular (green) and \regularreverse (shaded-green).
However, more \intractable cookies are sent in the \regular run compared to \regularreverse run on average (\ie the median in the red boxplot is one while it is zero for the shaded-red boxplot). This may indicate that either more popular websites set or the less popular ones send more \intractable cookies. We further investigate the impact of website rankings in~\Cref{subsec:ranking}. 
For the random runs (\Cref{fig:box_random}), the medians of \intractable cookies are zero for both, indicating relatively consistent behavior.

In total, it is evident that \intractable cookies are common across websites. In all runs, the majority of tracking cookies set in the \cookiejar are identified as \intractable, \eg out of 3,583 unique tracking cookies in the \regular run, 2,131 cookies are later classified as intractable (see~\Cref{app:cookie-jar} for details).
\todo{Note that in our measurements, as detailed in \Cref{subsec:classify}, \intractable cookies are by definition a subset of detected tracking cookies. Although we consider both first- and third-party cookies when identifying tracking cookies, $\approx95\%$ of them are third-party cookies in all runs.}

\begin{figure}[h!]   
    \includegraphics[width=0.95\columnwidth]{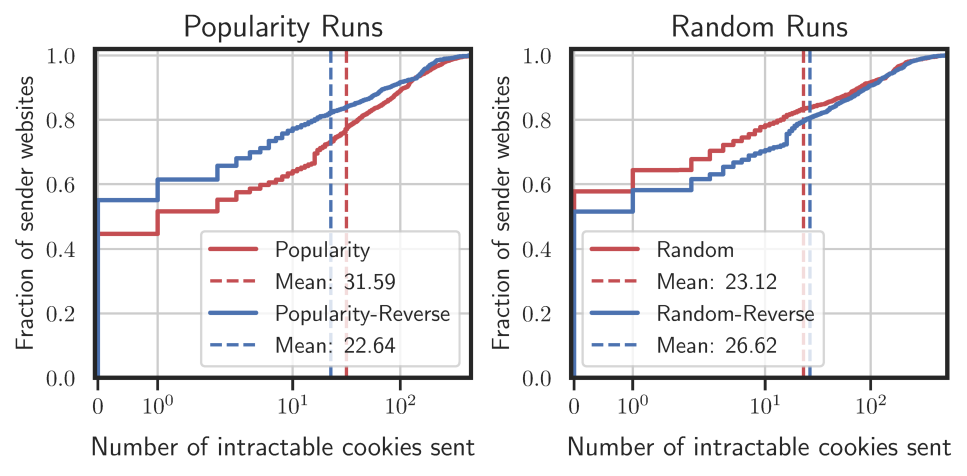}
    \caption{Intractable cookie distribution over websites.}    
    \label{fig:Intractable_cookie_ECDF}
\end{figure}

To further analyze the \intractable cookie distribution, we plot the Empirical Cumulative Distribution Function (ECDF) of \intractable cookies for popularity and random runs (see \Cref{fig:Intractable_cookie_ECDF}). The graphs indicate that around 45\% to 55\% of websites send at least one \intractable cookie across all runs.
Moreover, we observe on average $\approx40\%$ more \intractable cookies in the \regular run compared to the \regularreverse run, while this difference is relatively smaller for random runs, again indicating the possible impact of website ranking on \intractable cookies (see \Cref{subsec:ranking}).

\subsection{Impact of Banner Interaction}
\label{subsec:intract}

By definition, \intractable cookies are sent to tracker domains prior to any interaction with the banner. We now investigate the possible effect of banner interaction on the later transmission of \intractable cookies. As mentioned in \Cref{sec:meth}, for each domain in the stateless phase, BannerClick performs two separate iterations: one for rejecting the banner and then reloading the webpage, and one for accepting it.

\begin{figure}
    \centering
    \scalebox{0.915}{ %
    \begin{minipage}{\columnwidth}
     \begin{subfigure}[t!]{0.4255\columnwidth}
         \centering
         \vspace{0.54mm} 
         \includegraphics[width=\textwidth]{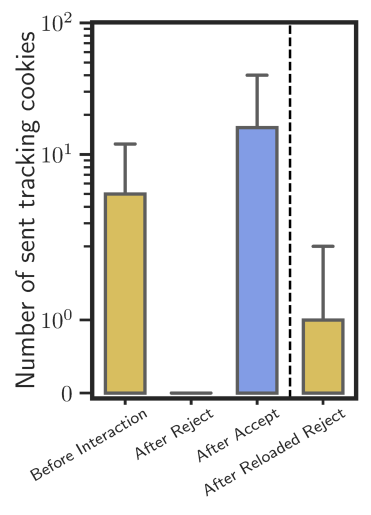}
         \captionsetup{skip=-1.8pt, font=scriptsize}
         \caption{Impact of Banner Interaction }
         \label{fig:interact_impact}
     \end{subfigure}
     \hfill
          \begin{subfigure}[t!]{0.558\columnwidth}
         \centering
         \vspace{0.45mm} 
         \includegraphics[width=\textwidth]{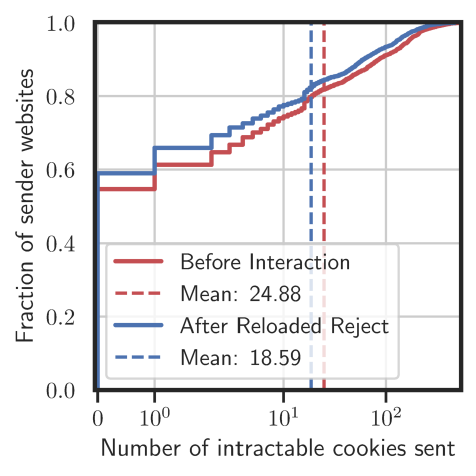}
         \captionsetup{font=scriptsize}
         \caption{Impact of Reload}
         \label{fig:reload_impact}
     \end{subfigure}
        
        \caption{Impact of banner interaction and reloading on the number of tracking cookies in Cookie Jar sent. Yellow boxes in (a) are \intractable cookies whose ECDF plot is shown in (b) for Before Interaction and After Reloaded Reject.}
        \label{fig:interact-reload_impact}
    \end{minipage}
    }
\end{figure}

The first three boxplots in \Cref{fig:interact_impact} show the number of tracking cookies sent upon visiting a webpage that were previously stored in the \cookiejar, corresponding to three stages: before interaction (\ie \intractable cookies), after rejecting, and after accepting the cookie banner for \randcomb run. The flat line for the ``After Reject'' box indicates that rejecting the banner does not trigger sending additional tracking cookies previously set. We also find that just a few cookies are explicitly deleted after rejecting the banners. In contrast, accepting the banner triggers the transmission of a new set of cookies (\ie the blue box), which can be considered a valid action since it reflects user consent to the banner. \todo{Additionally, we observe that cookies set after accepting the banner have a median expiration time of 6 months—twice as long as \intractable cookies set prior to banner interaction, which have a median of 3 months.}

Additionally, the right-most box in \Cref{fig:interact_impact} shows the number of \intractable cookies continuing to be sent after revisiting the rejected website (\ie a subset of \intractable cookies). Compared with the ``Before Interaction'' box, we see the third quartile dropping from 5 to 1. To further assess the impact of this reload, we plot the ECDF graph of \intractable cookies in \Cref{fig:reload_impact}, comparing the stages before interaction and after reloading the rejected webpage. We observe that, on average, websites stop sending $\approx25\%$ of \intractable cookies after reloading the rejected webpages. 

Overall, considering the ``After Accept'' stage, we observe that initiating tracking after user interaction with banners is technically feasible. However, as the prevalence of \intractable cookies indicates, prioritizing tracking before obtaining consent (\ie the ``Before Interaction'' stage) remains a common practice on the web.

\vspace{1.5mm}

\noindent\textbf{A Case Study:}  
Through our manual inspection, we observed that interacting with a banner can modify a website's front-end configuration. For instance, rejecting a banner may exclude third-party tracking resources from the HTML source. Conversely, accepting a banner may prompt websites to inject additional third-party resources into the HTML, leading to the transmission of new cookies.

Notably, we found that one of the major entities that govern the website's behavior regarding loading third-party resources is the Consent Management Platform (CMP) (see \Cref{subsec:banner} for more details).
For example, \texttt{ritzcarlton.com} uses \texttt{OneTrust}, one of the most popular CMP~\cite{rasaii2023exploring, hils2020measuring}, to manage user preferences and interactions with the banner. Upon visiting \texttt{ritzcarlton.com} for the first time, it renders an iFrame from \texttt{demdex.net}, a domain associated with Adobe Audience Manager.\footnote{Adobe Audience Manager is a Data Management Platform (DMP) that collects, manages, and segments user data for personalized advertising and audience targeting.}
In this case, the script inside the iFrame creates new cookies and sends them via XMLHttpRequest API calls to different tracker domains. By default, the browser also sends all previously set cookies along with these requests. Tracker domains can potentially link these cookies together and create a user profile using cookie synchronization techniques~\cite{syncCookies2019}. After the user rejects the banner, the CMP stores the user's preference in a cookie, stopping the iFrame from loading and preventing further requests to trackers on subsequent visits to the webpage.

In total, out of the 127,645 \intractable cookies detected in the \randcomb run, around 73\% are sent as a result of HTTP requests made by the browser to fetch third-party resources (\eg img, script, beacon, etc.). The remaining 27\% cookies are directly sent via XMLHttpRequest API calls (\eg \textit{fetch()} method) from script codes.

\subsection{Impact of Global Privacy Control}
\label{subsec:gpc}

In addition to interacting with cookie banners,
other official, standardized mechanisms have also been developed. One of the most notable is Global Privacy Control (GPC)~\cite{GlobalPr60:online}, which has recently gained more attention and is now supported by many browsers and extensions.\footnote{https://globalprivacycontrol.org/orgs}
GPC is a browser setting that signals a user's preference not to be tracked. When enabled, the browser informs websites that users do not want their data to be sold or shared. 
In 2021, the California Attorney General confirmed that businesses must honor the GPC signal as a valid request to opt out of data sales under CCPA~\cite{ccpa_gpc_faq}. While GDPR does not explicitly mandate GPC, the signal can be interpreted as a withdrawal of consent for tracking, which websites operating under GDPR should honor.

\begin{figure}[t!]
\centering
\vspace{0.460mm} 
\includegraphics[width=0.51\columnwidth]{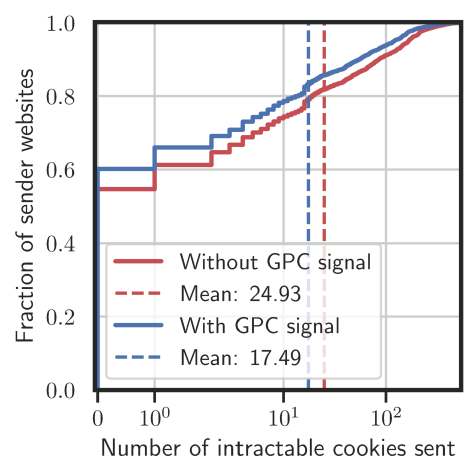}
\caption{Impact of enabling GPC signal on the number of \intractable cookies.}
\label{fig:gpc_fig}
\vspace{-2mm}
\end{figure}

To measure its impact on the \intractable cookies, we perform another stateless run as detailed in \Cref{meth:collection} with the GPC signal enabled. \Cref{fig:gpc_fig} shows the difference in \intractable cookies between the \randcomb run with and without GPC enabled. We observe that enabling GPC reduces \intractable cookies by approximately 30\% on average, and about 68\% of the remaining cookies overlap with those observed in the ``After Reloaded Reject'' case shown in \Cref{fig:interact-reload_impact}.
This indicates that enabling the GPC signal reduces \intractable cookies by 30\%, and further rejection of the banner can lead to an additional 32\% reduction on subsequent visits, resulting in an average of approximately 11.9 intractable cookies per domain.

\subsection{Impact of Website Rank}
\label{subsec:ranking}

\begin{figure}[t!]
\centering
\vspace{1.0mm}
\includegraphics[width=0.464\columnwidth]{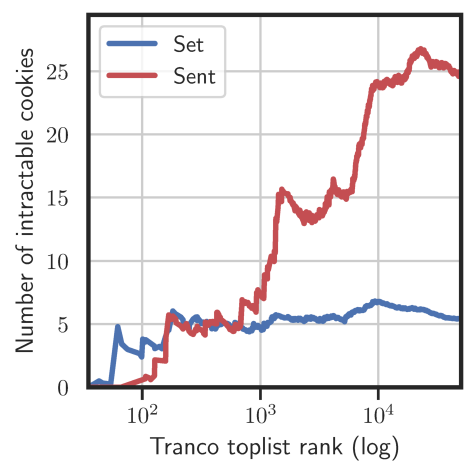}
\caption{Average number of set and sent \intractable cookies over Tranco's toplist ranks.}
\label{fig:rank_randcomb}
\end{figure}

\Cref{fig:rank_randcomb} depicts the average number of \intractable cookies set and sent by Tranco websites as per their rank tier.
For this analysis, we use the \randcomb run, which contains 5,915 accepted domains in the stateful phase and 5,121 rejected domains in the stateless phase. The red line shows the average number of \intractable cookies sent based on the top list rank of the rejected domains. Accordingly, the blue line shows the average number of set cookies over the top list rank of the accepted domains.

For the red line, we observe an ascending trend where popular websites, on average, send fewer \intractable cookies than less popular ones. For instance, we observe that the top 50 websites send, on average, zero \intractable cookies.\footnote{This observation is supported by our analysis of popularity runs, where none of the 6 rejected domains in the top 50 sent \intractable cookies.} 
Whereas the top 10k websites send, on average, about 25 \intractable cookies. This trend may be explained by the tendency of more popular websites to utilize their own resources, potentially resulting in fewer HTTP requests to third parties and, consequently, fewer \intractable cookies being sent.  
Conversely, the blue line remains constant, with websites setting around 5 \intractable cookies on average, regardless of their relative ranking.

\subsection{Impact of Banner Type}
\label{subsec:banner}

\begin{figure}[t!]
     \centering
    \scalebox{0.949}{ %
    \begin{minipage}{\columnwidth}
     \begin{subfigure}[t!]{0.470\columnwidth}
         \centering
         \vspace{0.35mm} 
         \includegraphics[width=1.055\textwidth]{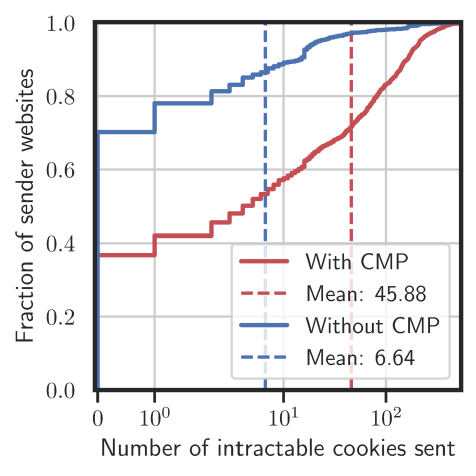}
         \caption{Impact of CMP}
         \label{fig:cmp_impact}
     \end{subfigure}
     \hfill
          \begin{subfigure}[h!]{0.513\columnwidth}
         \centering
         \vspace{0.60mm} 
         \includegraphics[width=1.06\textwidth]{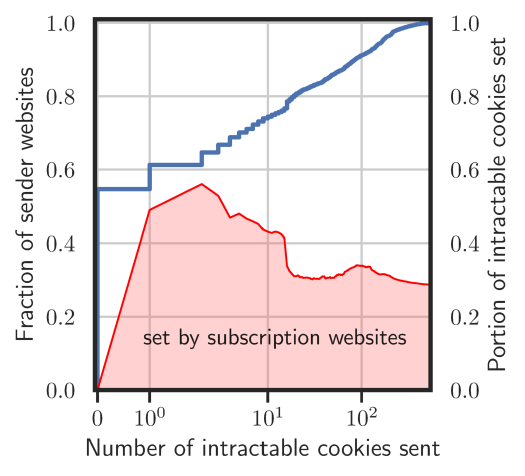}
         
         \caption{Impact of Cookie Paywall}
         \label{fig:cookie_paywall}
     \end{subfigure}
        \caption{Impact of type of banners on the number of \intractable cookies for RandComb run.}
        \label{fig:type_of_banners}
     \end{minipage}}
\end{figure}

\noindent\textbf{CMP Banners:} 
As mentioned in \Cref{subsec:intract}, Consent Management Provid\-ers (CMPs) are another entity involved in the context of 
intractable cookies. Out of 5,121 rejected banners in the \randcomb run, BannerClick classifies 2,386 as CMP banners. \Cref{fig:cmp_impact}
illustrates the difference in the number of \intractable cookies between websites with and without CMPs. On average, websites embedding CMP banners set 6.91 times more \intractable cookies than those using native banners, highlighting a significant discrepancy in cookie deployment. We also find that CMP banners are relatively harder to reject, as $\approx40\%$  of them require exploring of ``Settings'' menu by \bannerclick, involving more than one click to be fully rejected. Whereas over 90\% of native banners have a direct ``Reject'' button and can be rejected with a single click.
\todo{Overall, our findings suggest that CMPs often do not prioritize facilitating user consent or ensuring strict compliance with privacy regulations. They are generally harder to reject and tend to transmit significantly more \intractable cookies than native banners.\footnote{These differences may be due to the likelihood that websites with complex ad interdependencies are more inclined to use CMPs and communicate with trackers.} This aligns with the origin of many CMPs in the IAB Europe Transparency and Consent Framework, which was developed by an industry organization representing the interests of the online advertising sector.}

\vspace{2mm}

\noindent\textbf{Cookie Paywalls:}
\todo{In contrast to typical cookie banners, another form---known as a ``cookie paywall''---offers more restricted options to users.}
Cookie paywalls require users to either opt in to banner policies (mostly tracking) or pay for an ad-free browsing experience through a subscription~\cite{rasaii2023thou}.
\Cref{fig:cookie_paywall} displays the ECDF graph for the \randcomb run, highlighting the portion of \intractable cookies set by websites that display cookie paywall-based banners (totaling 90 detected by \bannerclick out of 5,915 accepted websites). Note that, following the blue line, 60\% of websites send a maximum of 2 \intractable cookies, of which around 55\% of the cookies are set or reset by websites with cookie paywall banners (peak of the red area in the graph). Interestingly, this proportion drops to around 30\% and remains relatively constant for websites that send more than 20 \intractable cookies.
Overall, this shows that even if users somehow manage to reject all other banners, they still get substantial \intractable cookies from a few cookie paywalls.

\subsection{Characteristics of Intractable Cookies}
\label{subsec:cookie-exp-dup}

\noindent{\textbf{Expiration and Renewal Analysis:} 
As mentioned in \Cref{sec:meth}, we standardize the expiration times of all retrieved cookies to a uniform date far in the future (Saturday, 01 January 2028, 12:12:12 UTC) to increase the consistency of our measurements and enhance the reproducibility of the findings. We recognize that this approach of artificially extending cookie expiration times may introduce bias into our analysis.
Additionally, the diversity and the number of cookies users encounter can vary based on their individual browsing behaviors and the websites they visit. Thus, to better assess the validity of our results, we examine the actual expiration times of \intractable cookies in the \cookiejar, \todo{as well as the number of times they are set or reset across different accepted websites.}
More detailed analyses and plots are presented in \Cref{app:cookie-exp-dup}.

\Cref{fig:heatmap_reg_pair} illustrates the distribution of unique\footnote{Cookies are grouped by their \textbf{name} and \textbf{host} attributes.} \intractable cookies in the popularity runs, based on their expiration times and the number of websites that set them upon banner acceptance.
On the x-axis, expiry times are segmented into 1 day, 10 days, 3 months, and 1 year intervals, with ``Session'' included as a special category for session cookies.
These are the common expiry durations for cookies. For example, nearly 25\% of \intractable cookies have an expiry time set to 1 year (see \Cref{app:cookie-exp-dup}).
The y-axis represents buckets corresponding to the percentage of accepted websites setting \intractable cookies, including a distinct category for cookies set only once \todo{(\ie the first row labeled with the bolded 1)}.
The numbers displayed within the heatmap cells represent the count of \intractable cookies.
For instance, the bottom-right cell of the heatmap for the \regular run shows that 553 unique \intractable cookies have an expiry time exceeding one year and were set by a single website during the entire crawl. Both heatmaps show that about 90\% of cookies are set by no more than 1\% of websites. 
Furthermore, our analysis reveals that approximately 40\% of these cookies have expiration dates of at least one year, indicating a common practice among websites to track users over extended periods. Additionally, over 65\% are set with an expiration time of at least 10 days (see \Cref{app:cookie-exp-dup} for details), which aligns with the duration needed to complete the stateful phase and populate the \cookiejar across all four runs. Therefore, for more than 65\% of \intractable cookies, we can state that our adjustment of expiration times does not lead to an overestimation of their prevalence.   
However, we cannot draw conclusive insights regarding session cookies or those with shorter expiration durations. A similar distribution pattern is observed in the random runs.

\begin{figure}[t!]
\centering
\includegraphics[width=0.999\columnwidth]{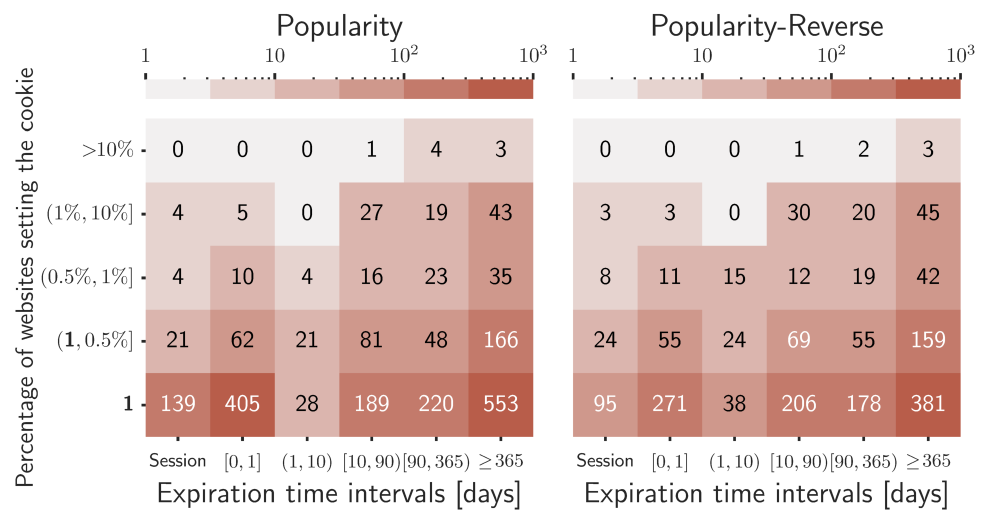}
\caption{Distribution of \intractable cookies based on the number of websites setting them and expiration time.
The first row represents a single website setting the cookies.
}
\label{fig:heatmap_reg_pair}
\vspace{-4mm}
\end{figure}

\vspace{1.1mm}

\noindent{\textbf{Cookie Synchronization Analysis:} \todo{Additionally, trackers can employ various techniques—such as cookie synchronization~\cite{bouhoula2024automated, 10.1145/3442381.3450056}—which enable them to link identifiers across domains and construct more comprehensive user profiles. Following the methodology used in prior works, we examine the transmission of cookie values via redirection URL parameters. We find that, out of 2,545 unique \intractable cookies in the \randcomb run, 76 (\ie 3\%) are synchronized at least once with other \host through redirection parameters.}
\todoo{Moreover, we manually inspect the values of the top 100 intractable cookies and find that they predominantly contain encrypted or encoded strings, likely serving as unique user identifiers or as part of mechanisms for continued tracking. A smaller portion of cookies contain simple values such as numbers or binary flags (e.g., ``YES'', ``true''), which appear to store preferences or session states. For our cookie syncing detection, we exclude these simpler cases and focus only on cookies containing encrypted or encoded strings longer than 10 characters.}

\definecolor{headerColor}{RGB}{79,129,189} %
\definecolor{rowColor}{RGB}{221,235,247} %
\definecolor{rowColorEven}{RGB}{200,235,200} %
\definecolor{amber}{rgb}{1.0, 0.75, 0.0}
\definecolor{amber(sae/ece)}{rgb}{1.0, 0.49, 0.0}
\definecolor{cream}{rgb}{1.0, 1.0, 0.90}
\definecolor{bisque}{rgb}{1.0, 0.89, 0.77}
\definecolor{almond}{rgb}{0.94, 0.87, 0.8}
\definecolor{burlywood}{rgb}{0.73, 0.58, 0.40}

\newlength{\mylength}
\setlength{\mylength}{10.5pt}

\setlength{\heavyrulewidth}{1.5pt}  %
\setlength{\lightrulewidth}{0.0001pt} %
\setlength{\cmidrulewidth}{1.7pt}  %
\setlength{\extrarowheight}{1.3pt}
\setlength{\aboverulesep}{0pt}
\setlength{\belowrulesep}{0pt}

\begin{table*}[t!]
\small
\centering
\resizebox{0.999\textwidth}{!}{%
\begin{tabular}{c@{\hspace{10.5pt}}c@{\hspace{10.5pt}}c@{\hspace{10.5pt}}c@{\hspace{10.5pt}}c@{\hspace{11.2pt}}c@{\hspace{10.5pt}}c@{\hspace{10.5pt}}c@{\hspace{10.5pt}}c@{\hspace{10.5pt}}c}
\multicolumn{10}{c}{} \\
\rowcolor{burlywood} 
 \cellcolor{white} & \color{white}\textbf{Tracker Domain} & \color{white}\textbf{\#Cookies} & \color{white}\textbf{\#UniqCookies} & \color{white}\textbf{\#Senders} & \cellcolor{white} & \color{white}\textbf{Tracker Domain} & \color{white}\textbf{\#Cookies} & \color{white}\textbf{\#UniqCookies} & \color{white}\textbf{\#Senders} \\
\midrule
\rowcolor{almond}
1 & amazon-adsystem.com & 1,526 & 2 & 763 & 11 & 3lift.com & 1,240 & 5 & 365 \\
\rowcolor{cream}
2 & adsrvr.org & 1,344 & 2 & 672 & 12 & lijit.com & 1,084 & 13 & 343 \\
\rowcolor{almond}
3 & criteo.com & 2,623 & 7 & 649 & 13 & bidswitch.net & 2,723 & 11 & 323 \\
\rowcolor{cream}
4 & pubmatic.com & 19,905 & 40 & 621 & 14 & a-mo.net & 5,458 & 18 & 311 \\
\rowcolor{almond}
5 & adnxs.com & 3,020 & 5 & 604 & 15 & taboola.com & 6,118 & 42 & 293 \\
\rowcolor{cream}
6 & casalemedia.com & 2,128 & 4 & 532 & 16 & nr-data.net & 282 & 1 & 282 \\
\rowcolor{almond}
7 & id5-sync.com & 1,054 & 7 & 517 & 17 & bidr.io & 414 & 2 & 282 \\
\rowcolor{cream}
8 & openx.net & 1,203 & 3 & 401 & 18 & tapad.com & 771 & 3 & 257 \\
\rowcolor{almond}
9 & smartadserver.com & 2,074 & 7 & 375 & 19 & liadm.com & 444 & 2 & 243 \\
\rowcolor{cream}
10 & sharethrough.com & 2,357 & 12 & 366 & 20 & quantserve.com & 454 & 2 & 227 \\
\noalign{\vskip 1.1pt}
\bottomrule
\end{tabular}
}
\captionsetup{justification=centering, skip=0pt, width=0.8\textwidth}
\caption{Top 20 \hosts with the total and unique number of \intractable cookies associated with them sorted by the \senders sending them.}
\label{table:toptrackers}
\end{table*}

\vspace{-1mm}

\subsection{Domain Analysis}
\label{subsec:domain-analysis}

In this section, we analyze the roles and prevalence of \sender and \host in the transmission of \intractable cookies.

\begin{figure}[t!]
\centering
\includegraphics[width=0.52\columnwidth]{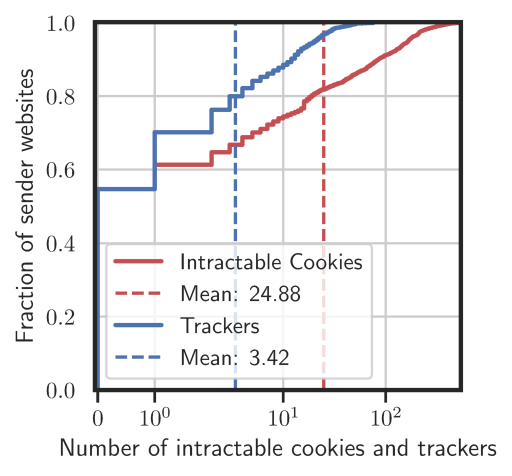}
\caption{Comparison of the number of \intractable cookies and the number of associated trackers per \sender.}
\label{fig:host_cookie}
\end{figure}

The ECDF plots in \Cref{fig:host_cookie} compare the number of \intractable cookies and the number of \hosts
(responsible for eliciting them) corresponding to the \textit{sender websites} for the \randcomb run. On average, each \sender is associated with 3.42 different \hosts, each with at least one \intractable cookie. Furthermore, based on the means, a single \sender has an average of 7.3 \intractable cookies per \host.

In \Cref{table:toptrackers}, we present the top 20 \hosts along with the total and unique number of \intractable cookies associated with them, sorted by the number of \senders responsible for dispatching them.  
We observe that most top trackers use only a handful of unique cookies to track users across hundreds of the 5,121 rejected websites.
We manually verified that the majority of the top 50 domains owning \intractable cookies belong to recognized ad tech companies specializing in programmatic advertising solutions (\eg \texttt{pubmatic.com}), while others provide analytic tools and feedback through session recordings and surveys (\eg \texttt{hotjar.com}).

\subsection{Partitioned Cookies Analysis}
\label{subsec:partitioned}

\todo{Partitioned cookies are a newly proposed privacy feature in Chrome \cite{CookiesH71:online}, also adopted by browsers like Firefox. They scope third-party cookies to the top-level site, preventing them from being shared across different websites.
Accordingly, they have the potential to mitigate \intractable cookies by isolating their transmission to each \setter.}
For example, in the scenario described in \Cref{subsec:intract}, when \texttt{basic.com} sets a cookie from \texttt{tracker.net}, the cookie is stored in a partitioned context specific to \texttt{basic.com}. Later, when the user visits \texttt{new.com} (\sender), even if it loads a resource from \texttt{tracker.net}, the browser does not send the previously set cookie ($id=123$) because partitioned storage treats \texttt{new.com} as a separate context. As a result, \texttt{tracker.net} cannot correlate the user’s activity across \texttt{basic.com} and \texttt{new.com}.

\begin{table}[t!]
    \centering
    \resizebox{0.999\linewidth}{!}{
    \begin{tabular}{lrrr}
    \toprule
    \textbf{Cookie Type} & \textbf{Total Unique} & \textbf{Partitioned} & \textbf{Along with NP} \\
    \midrule
    All Cookies        & 79,898  & 521  & -  \\
    Tracking Cookies   & 3,177   & 40   & 26   \\
    \bottomrule
    \end{tabular}
    }   
    \caption{Summary of unique partitioned cookies and those accompanying the non-partitioned (NP) cookies
    from the same \host.
    }
    \label{tab:partitioned_summary}
    \vspace{-10mm}
\end{table}

As detailed in \Cref{meth:collection}, we conduct a separate run using Chrome to measure the prevalence of partitioned cookies. The results reveal a distribution of cookie types similar to the other runs conducted with Firefox. As shown in \Cref{tab:partitioned_summary}, out of 79,898 unique stored cookies, 521 are marked partitioned.
However, among 3,177 unique tracking cookies, only 40 (1.3\%) are partitioned, of which 26 are accompanied by non-partitioned tracking cookies from the same tracker, with 9 having the same value.\footnote{In most cases, the partitioned cookie has a similar \texttt{name}, appended by 'p' or '\_p'.}
Interestingly, we also observe that Chrome does not overwrite existing cookies when their partitioned attribute differs. In other words, the same cookie can be set twice—once with the partitioned attribute and once without—both of which are sent in subsequent HTTP requests.

Overall, despite nearly three years since Google introduced partitioned cookies through the Cookies Having Independent Partitioned State (CHIPS) initiative with the release of Chrome 100 in March 2022, their adoption remains limited and gradual. This finding aligns with a recent study on CHIPS~\cite{CHIPSpam2025}. Nevertheless, a longitudinal study is needed to assess whether partitioned cookies will achieve widespread adoption and be effectively applied in real-world scenarios. We discuss partitioned cookies further in \Cref{sec:discuss}.
\section{DISCUSSION}
\label{sec:discuss}

In this section, we examine the relationship between banner designs, \intractable cookies, and privacy regulations, exploring mitigation strategies such as partitioned cookies and their limitations. We then introduce a Browser-Integrated Consent Mechanism as our proposed approach to addressing the shortcomings of the current consent mechanism that lead to \intractable cookies.

\subsection{Interpretation of Privacy Regulations}

\label{conclusion:gdpr}

\todo{The existence of \intractable cookies can be traced back to the impractical implementation of consent mechanisms, primarily in the form of cookie banners, introduced in response to the ePrivacy Directive and shaped further by the consent requirements defined under the GDPR. In the following, we examine the current deployment of cookie banners by addressing two key limitations in the broader privacy framework: the fragmented interpretation of the valid consent requirements and the ambiguity regarding which entities are responsible for ensuring compliance.}

\vspace{1.5mm}

\parax{Fragmented Interpretation of Consent Requirements:} 
\todo{While the GDPR outlines core criteria for valid consent, such as being freely given, specific, informed, and unambiguous, it does not prescribe how these requirements should be technically implemented. Instead, practical interpretation has been shaped through soft law instruments, including guidelines from the European Data Protection Board (EDPB)~\cite{edpb_about} and national Data Protection Authorities (DPAs)~\cite{bfdi2023cookiebanner},\footnote{\todo{The European Data Protection Board (EDPB) is an independent EU body that ensures consistent application of the GDPR across member states. On the other hand, National Data Protection Authorities (DPAs) are country-level regulators responsible for enforcing data protection laws and issuing context-specific guidance.}} as well as court rulings.\footnote{\todo{For example, the ECJ’s \textit{Planet49} decision explicitly ruled that pre-ticked checkboxes do not constitute valid consent~\cite{ecj2019planet49}.}}}

\todo{Although the EDPB uses the GDPR as its legal basis to harmonize data protection practices across the EU and to interpret vague or open-textured provisions, individual DPAs retain discretion in how these interpretations are applied and enforced at the national level. This can lead to divergent regulatory outcomes among Member States. For instance, the EDPB's Cookie Banner Taskforce Report (2023)~\cite{edpb2023cookiebanner} notes that most DPAs consider the absence of a ``Reject'' option on any layer where a consent (“Accept”) button is present to be non-compliant with the ePrivacy Directive. However, some authorities, such as the Spanish Data Protection Authority (AEPD), have adopted a more permissive stance, allowing the ``Reject All'' option to appear only in a secondary layer. This regulatory divergence contributes to a fragmented enforcement landscape. In practice, recent studies report widespread non-compliance and inconsistent implementation of cookie banners across websites~\cite{santos2020cookie, Matte2020respect}.}

\todo{Subsequently, as mentioned in \Cref{meth:bannerclick}, many cookie banners place the ``Reject All'' option within ``Settings'' layers, requiring users to navigate through granular settings often spread across multiple tabs. This design introduces unnecessary friction and undermines the effectiveness of cookie banners in clearly communicating data collection practices~\cite{wehner2023you, utz2019informed, jha2023refuse, mehrnezhad2021can}. Even when granular choices are necessary, they could be presented in a more user-friendly manner—accessible yet unobtrusive to the majority of users.}

\todo{Moreover, although the GDPR provides a general definition of valid consent, terms such as “freely given” and “unambiguous” remain open to interpretation. Prior research shows that cookie banners frequently employ deceptive patterns that nudge users toward acceptance through habituation rather than informed choice~\cite{soe2020circumvention, borberg2022so, 10.1145/3313831.3376321, coventry2016personality}. The rise of cookie paywalls~\cite{rasaii2023thou} further pressures users into accepting tracking, often against their preferences. As a result, many users reluctantly consent to tracking on certain sites, with little to no ability to revoke that consent later~\cite{mehrnezhad2021can}. As we showed, many of these seemingly isolated acceptances propagate across visits via \intractable cookies, enabling inter-domain tracking even before users provide explicit consent on the visited site.}

\vspace{1.5mm}

\parax{Ambiguity of Accountability:} 
\todo{The transmission of \intractable cookies to \hosts raises concerns under existing privacy regulations. The GDPR requires valid user consent before collecting or processing personal data, unless another legal basis applies—such as necessity for a service explicitly requested by the user (Article 6).
While the GDPR designates the ``data controller''\footnote{Article 4(7) GDPR defines the \textit{data controller} as ``the natural or legal person, public authority, agency, or other body which, alone or jointly with others, determines the purposes and means of the processing of personal data''.} as responsible for ensuring compliance (Article 24), accountability becomes complex when multiple parties are involved. Article 26 introduces the concept of ``joint controllers'' and requires an \emph{arrangement} to define their respective responsibilities. }

\todo{According to the EDPB Guidelines~\cite{edpb2021controller}, joint controllership arises when two or more entities jointly determine the purposes and means of data processing, regardless of whether they have equal access to the data. This was affirmed by the CJEU in the \textit{Fashion ID} ruling,\footnote{In \textit{Fashion ID} (C-40/17), the Court held that a website embedding a third-party plugin (e.g., Facebook's Like button) could be a joint controller, even without accessing the collected data, if it contributes to the determination of purposes and means.} establishing that shared decision-making alone can trigger shared responsibility.
In practice, however, applying this principle is challenging within the web's complex tracking ecosystems, involving multiple actors such as websites, third-party trackers, advertisers, DMPs, and CMPs. These actors rarely establish or disclose clear joint controller arrangements, making it difficult to determine who is accountable for informing users or fulfilling their data rights. This lack of clarity hinders enforcement and risks leaving users without a clear path to contest or revoke consent, ultimately limiting the GDPR's effectiveness in protecting personal data in such distributed environments.}

\vspace{1mm}

\todo{In a nutshell, the current deployment of cookie banners often reflects ad hoc compliance efforts, aimed more at fulfilling legal formalities than enabling meaningful user choice, creating a potential \textit{false sense of privacy}. While privacy regulations
promote transparency and accountability, their effectiveness hinges on a clear understanding of user behavior and the roles of various actors in the consent ecosystem. Without coordinated input from regulators, developers, and interdisciplinary experts, these frameworks risk undermining the very privacy they seek to protect by inconsistent enforcement and superficial implementation.}

\subsection{Intractable Cookies Mitigation} 
\label{conclusion:mit}

As discussed in \Cref{subsec:intract}, from a technical perspective, the core cause of \intractable cookies may be the \senders' inability to determine the existence of cookies set by previously visited websites (\ie \setters). 
Consequently, several solutions can mitigate or eliminate the \intractable cookies phenomenon. One approach is to \emph{prevent the loading of third-party resources} before banner acceptance. However, given the current structure of consent mechanisms---where websites handle user preferences via banners and browsers control request transmission---this solution is not feasible.

Alternatively
\emph{blocking third-party cookies} entirely is the most straightforward approach to mitigate the privacy-intrusive nature of tracking cookies, including \intractable cookies. This approach has already been implemented by browsers like Safari as a default setting~\cite{fingas2020safari}.
As for Chrome, which accounts for over 65\% of the browser market share across both desktop and mobile platforms~\cite{statcounter2025browser}, privacy-conscious users have the option to customize their settings and block third-party cookies.
However, studies~\cite{mehrnezhad2021can} show that most users are unaware of these controls or the tracking technologies behind them.  
More importantly, the debate over tracking extends beyond individual user preferences, as it involves conflicting interests among users, publishers, and advertisers. These competing priorities complicate the feasibility of outright blocking. For instance, Google's July 2024 reversal \cite{picchi2024google} of its 2020 pledge \cite{google_ending, patel2020google} to phase out third-party cookies in Chrome underscores the tension between privacy advocacy and economic interests.

Nonetheless, if we focus solely on users' interests and assume that blocking third-party cookies enhances their web experience by improving user privacy, the reality is more complex. 
Entities dependent on advertising revenue will likely shift to alternative tracking methods, such as fingerprinting~\cite{Iqbal2021Fingerprinting, Angove-Plumb2022Browser}, or adjust their pricing strategies to compensate for the loss of targeted ads~\cite{miller2023economicconsequencesonlinetracking}. Consequently, from a broader perspective, eliminating third-party cookies entirely may not provide the anticipated benefits for users unless an alternative monetization model is introduced.

Another possible solution to mitigate \intractable cookies is \emph{partitioned cookies}.  
However, besides its limited implementation (see \Cref{subsec:partitioned}), partitioned cookies also have several other limitations:

\begin{itemize}
    \item \emph{Developer Reliance:} Approaches that depend on widespread developer adoption often fail to achieve satisfactory results. For instance, studies on Content Security Policy (CSP)~\cite{Kerschbaumer2016InjectingCF, ubt_eref91468} reveal that less than 2\% of websites correctly implement it, with most deployments being ineffective or poorly configured. Although implementing partitioned cookies is less complex than CSP, it still requires modifying attributes like \texttt{SameSite} and appending the \texttt{\_Host} prefix to handle subdomains.

    \item \emph{Lack of Incentives:} 
    The incentive for adopting partitioned cookies remains unclear (particularly when considering the \host as the owner of the cookies) unless explicitly enforced by regulations. In comparison, CSP adoption is driven by its direct relation to the website's own security.
    \item \emph{Incompatibility with Consent-Based Tracking:} Most importantly, partitioned cookies lack the technical capability to enable inter-domain tracking, even upon explicit users' consent. This limitation undermines consent-based tracking, as it prevents users from selectively allowing or blocking tracking based on their preferences via banners.  
    For instance, some users may wish to receive personalized advertising while blocking tracking from specific websites, such as those handling sensitive content. However, with partitioned cookies, such granular control is no longer possible, rendering cookie banners ineffective for managing tracking preferences.
\end{itemize}

\subsection{Browser-Integrated Consent Mechanism} 

As discussed in \Cref{conclusion:gdpr}, one of the major drawbacks of the GDPR is its ambiguity in defining the ``data controller'' as the entity responsible for collecting and handling user consent. Moreover, obtaining user consent is distinct from data collection and processing and could instead be managed by a separate entity within the data flow.  
Accordingly, given the current structure of the web ecosystem---where the browser serves as the central element orchestrating communication between various entities---a practical approach is to integrate the consent mechanism directly into the browser. This would streamline the consent process, ensuring a more consistent and efficient implementation of consent mechanisms that can be easily scaled and updated.

In this model, browsers serve as the intermediary entity responsible for collecting user consent and preferences and applying them accordingly. This approach reduces the burden on both users and publishers (\ie developers). Users can configure their preferences through a well-structured consent management portal within the browser, ensuring their choices are consistently applied across all websites they visit. Likewise, publishers would no longer need to implement their own consent banners, resulting in a more uniform and user-friendly design.

\begin{figure}[t]
    \centering
    \includegraphics[width=0.85\columnwidth]{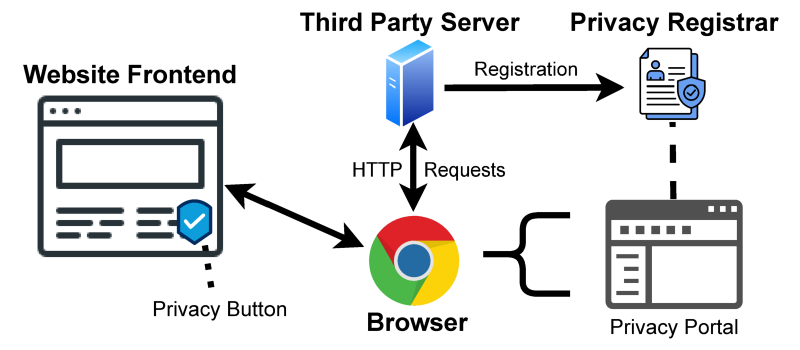}
    \caption{Overview of the proposed Browser-Integrated Consent Mechanism, consisting of four major components: browser, website, third party, and registrar.}
    \label{fig:bic}
\end{figure}

\todo{\Cref{fig:bic} depicts an overview of our proposed browser-integrated consent mechanism, which consists of four main components: the browser, the website, third parties (e.g., trackers or advertisers), and a privacy registrar. In this scheme, each third party intending to engage in cross-site tracking (e.g., via cookies) must register with a centralized registrar and disclose its intended purposes, similar in spirit to how vendors register with the IAB Europe Global Vendor List in the context of CMPs~\cite{tcf2}. However, unlike the IAB, which primarily represents the online advertising industry, the registrar in our proposal requires a more neutral and trustworthy governance model. Possible implementations include a government-backed registrar or a multi-stakeholder consortium similar to those organized by W3C~\cite{w3cGovernance2022}. The browser retrieves declarations from this registrar and uses them to decide, based on the user’s privacy preferences, whether to allow or block specific third-party requests.}

\todo{Users interact with consent controls through a unified, browser-operated system called the \emph{privacy portal}, which offers a consistent interface for managing preferences per website and registered third parties. The portal can be accessed through two entry points: (1) directly via the browser’s settings, and (2) through a \emph{privacy button} displayed in the corner of the viewport when visiting a website, which opens the relevant section of the portal for that site. For example, users can opt out of specific trackers or websites based on category, or adjust their settings at any time using the privacy button.
The interface’s structure and appearance remain consistent across all websites. Within the portal, visible “Accept All” and “Reject All” buttons are always available, alongside more granular controls populated using data retrieved from the privacy registrar.}

This high-level prototype can be further developed and refined to strengthen user privacy while maintaining essential advertising capabilities. 
One potential approach is for browsers to treat all third-party cookies as partitioned by default and transmit them cross-site only if explicitly permitted by user preferences.
Furthermore, the portal could incorporate a subscription-based model, allowing websites to monetize their content directly as an alternative to ad-supported tracking, similar to existing cookie paywalls. In this model, websites could adapt their behavior and render the front end based on user choices or subscription status.  
Although this model may create a divide between paying and non-paying users, its alignment with free market principles could, in the long run, lead to improved services and a more sustainable online environment.

\todo{Lastly, the browser-integrated approach aims to reduce complexity by consolidating consent interactions and preference management into a single, user-centric system. Unlike prior browser-level mechanisms such as Global Privacy Control (GPC) and Advanced Data Protection Control (ADPC)~\cite{human2022adpc}, our proposal operates independently of publisher support, \todoo{as browsers can now allow or block third-party requests based on the user's privacy preferences set via the privacy portal and information retrieved from the registrar.} This independence makes it more practical and enforceable in real-world scenarios, particularly when dealing with non-compliant or uncooperative websites. The primary challenge, however, lies in persuading or requiring browser vendors to adopt such a mechanism, an objective that could be facilitated through regulatory mandates or privacy-focused legislation.
}

\vspace{-2mm}

\section{CONCLUSION}

In this paper, we reveal the prevalence of \textit{\intractable} cookies---tracking cookies that are set by websites where users accept their banners, persistent in the user browser, and sent to tracking domains before the user's explicit consent on subsequent websites. Through extensive measurements involving 20,000 domains from the Tranco top list, we demonstrated that around 50\% of the websites sent at least one \intractable cookie. 
Furthermore, we assessed how banner interaction, enabling GPC signal, can contribute to preventing intractable cookies. We then explored the impact of the website rank and type of the banner on the prevalence of these cookies. Moreover, we analyzed the expiration and renewal characteristics of \intractable cookies, along with their domain distribution. Finally, we examined current technical solutions, such as partitioned cookies, that aim to mitigate \intractable cookies and discussed their limitations.

Overall, our findings highlight a gap between the technical and legislative aspects of the web tracking ecosystem, leading to solutions like cookie banners that add complexity without effectively addressing the core issue. We advocate for meaningful improvements through the conscious collaboration of all stakeholders, including developers, regulators, publishers, and advertisers.

\begin{acks}

We sincerely thank the anonymous reviewers for their thoughtful feedback, which significantly improved the readability and clarity of the paper, and acknowledge the Max Planck Institute for Informatics for funding the work and providing the infrastructure needed for the measurements and analysis.   
\end{acks}

\bibliographystyle{ACM-Reference-Format}
\bibliography{main}

\appendix

\newpage

\section{Ethical Considerations}

In conducting our measurements, we abide by the ethical guidelines proposed by Partridge and Allman \cite{partridge2016ethical} and Kenneally and Dittrich \cite{kenneally2012menlo}, and follow the best measurement practices as described by Durumeric et al. \cite{durumeric2013zmap}. Our methodology involves running \openwpm in an automated manner to crawl each website using a standard web browser configuration.
\todo{We utilize dedicated measurement machines configured with informative reverse DNS (rDNS) names, which are subdomains of a research-affiliated organizational domain and are easily identifiable.} Moreover, we offer stakeholders the option to opt out and be excluded from our measurements. Throughout our measurement period, we did not receive any complaints.

\section{Banner Screenshot}
\label{app:banners}

\Cref{fig:banner-preselect} shows an example of a cookie banner that provides granular choices for different partners and purposes after the user clicks the ``Settings'' button. In this case, all non-essential cookies are turned off by default. As a result, clicking the ``SAVE \& EXIT'' button should functionally correspond to rejecting the banner.

\begin{figure}[t!]
\centering
\includegraphics[width=0.94\columnwidth]{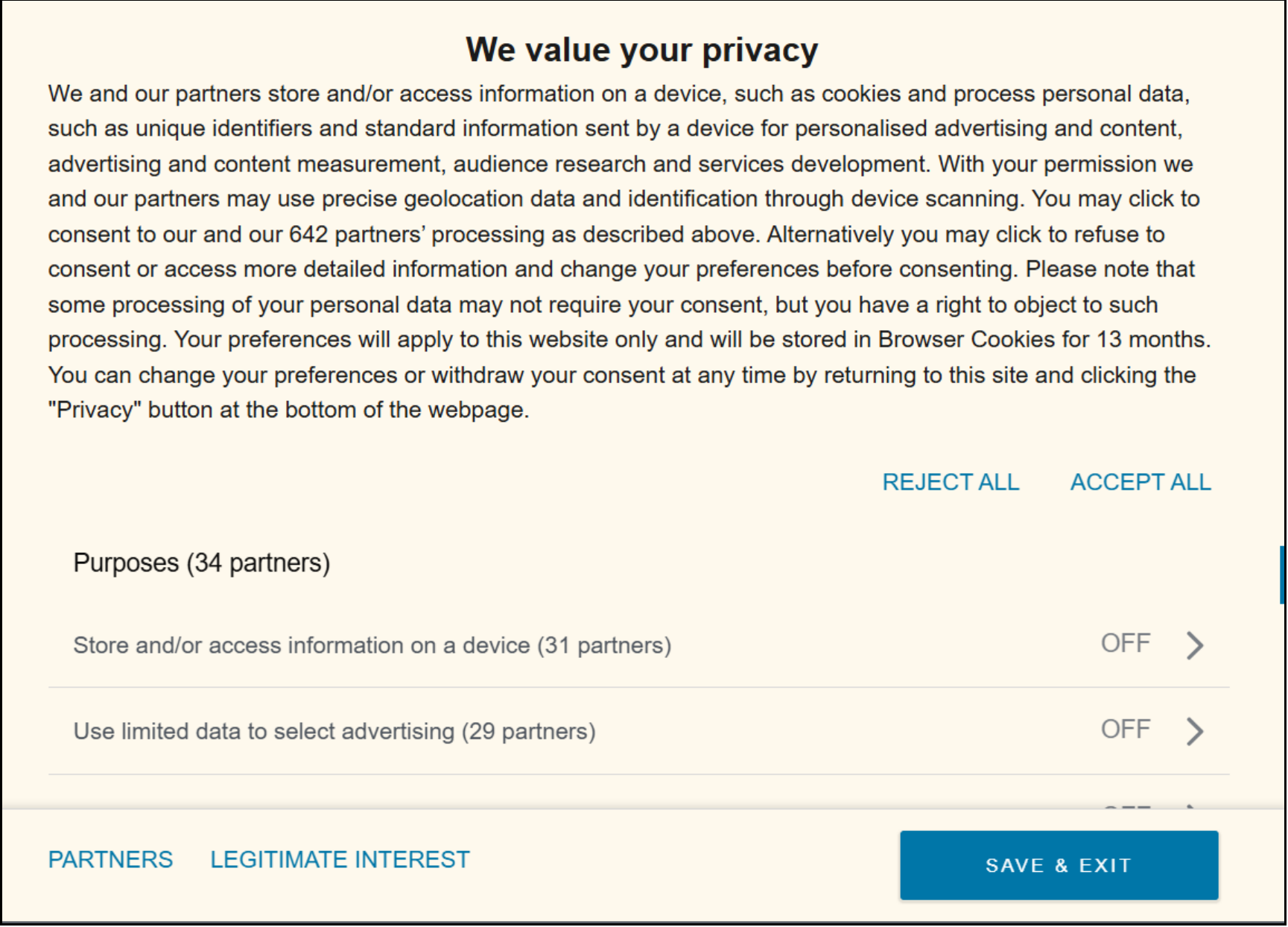}
\caption{Example of a banner displaying granular options after clicking the ``Settings'' button. Clicking ``SAVE \& EXIT'' should be equivalent to rejecting tracking, as all non-essential options are disabled by default.}
\label{fig:banner-preselect}
\end{figure}

\section{Cookie Jar and Sent Cookies}
\label{app:cookie-jar}

This section details the overall distribution of cookies collected during our measurement campaign across both popularity and random runs, as quantified in the numbers presented in \Cref{table:cookies}.

Overall, the table classifies cookies under two main categories (databases): \cookiejar and \sentcookies.
\cookiejar refers to cookies that are set by websites in the stateful phase when their consent banners are accepted by the BannerClick 
tool \cite{rasaii2023exploring}. \emph{Sent Cookies} includes cookies extracted from the HTTP requests of websites before rejecting consent banners. 
Subsequently, the table's sub-columns categorize the cookies into different types, viz., `Total,' `First Party,' `Third Party,' `Tracking,' `Intractable,` and `Reset.' Specifically, `Total' denotes the number of all cookies in the databases. `Reset' refers to intractable cookies overwritten within the domain sending it, \eg in \Cref{fig:schematic}, \texttt{new.com} may also reset the \intractable cookie ($id{=}123$).

For each run, the first row presents the cumulative count of cookies collected, reflecting the aggregated number of cookies set or sent across the domains. For instance, if a cookie is set twice by two different \setters, it is counted as two. On the other hand, the second row shows the count of unique cookies grouped by their \texttt{name} and \texttt{host} attributes. Finally, the third row, labeled `Avg,' depicts the mean number of cookies per domain.

Note that the table presents raw statistics without any adjustments. Across different runs, the number of accepted websites varies, influencing the number of cookies stored in the \cookiejar. Runs with a greater number of accepted domains are likely to have a larger number of unique cookies in \cookiejar, potentially resulting in a higher count of cookies sent per rejected domain. Consequently, a fair comparison cannot be made across the average numbers presented in the \sentcookies columns of different runs.
For a fair comparison, see \Cref{sec:results}.

In comparing \cookiejar of popularity runs, we observe that the average number of cookies is similar across all categories. 
In the case of random runs, the behavior across all categories appears relatively consistent. Additionally, in all runs, the counts of `Unique' \intractable cookies are identical between \cookiejar and its corresponding \sentcookies. This consistency is expected, as we classify a cookie in \cookiejar as \intractable only if it is subsequently sent from domains where consent is rejected.

Moreover, the average number of 'Reset' cookies is relatively low compared to the number of \intractable cookies sent. This highlights the cohort nature of \intractable cookies, as they tend to be sent via HTTP requests without being reset, making them difficult to track by simply observing the current state of cookies in the browser. It also implies that merely considering the deployment of tracking cookies overlooks a large portion of real-world tracking practices.

\definecolor{headerColor}{RGB}{79,129,189} %
\definecolor{rowColor}{RGB}{221,235,247} %
\definecolor{rowColorEven}{RGB}{200,235,200} %
\definecolor{amber}{rgb}{1.0, 0.75, 0.0}
\definecolor{amber(sae/ece)}{rgb}{1.0, 0.49, 0.0}
\definecolor{cream}{rgb}{1.0, 1.0, 0.90}
\definecolor{bisque}{rgb}{1.0, 0.89, 0.77}
\definecolor{almond}{rgb}{0.94, 0.87, 0.8}
\definecolor{burlywood}{rgb}{0.73, 0.58, 0.40}

\setlength{\mylength}{10.5pt}

\setlength{\heavyrulewidth}{1.5pt}  %
\setlength{\lightrulewidth}{0.0001pt} %
\setlength{\cmidrulewidth}{1.7pt}  %
\setlength{\extrarowheight}{1.3pt}
\setlength{\aboverulesep}{0pt}
\setlength{\belowrulesep}{0pt}
\begin{table*}[t]
  \small
  \setlength{\tabcolsep}{5pt} 
  \begin{tabular}{c@{\hspace{11.5pt}}c@{\hspace{11.5pt}}c@{\hspace{\mylength}}c@{\hspace{\mylength}}c@{\hspace{\mylength}}c@{\hspace{\mylength}}c@{\hspace{11.5pt}}c@{\hspace{\mylength}}c@{\hspace{\mylength}}c}
  & \multicolumn{6}{c} {\hspace{25pt} 
 \cellcolor{white}\color{black}\textbf{Cookie Jar}} & \multicolumn{3}{c}{\cellcolor{white}\color{black}\textbf{Sent Cookies}} \\ 

  \cmidrule(l{1pt}r{10pt}){3-7} \cmidrule(l{0pt}r{10pt}){8-10}
  \noalign{\vskip 0.2pt}
  \rowcolor{burlywood} 
    \cellcolor{white}&  \cellcolor{white}& \color{white}\textbf{Total} & \color{white}\textbf{First Party} & \color{white}\textbf{Third Party} & \color{white}\textbf{Tracking} & \color{white}\textbf{Intractable} & \color{white}\textbf{Total} & \color{white}\textbf{Intractable} & \color{white}\textbf{Reset} \\ [1.1pt]
  \midrule

  \rowcolor{almond}
  & Aggregate & 95,131 & 57,918 & 37,213 & 21,920 & 19,296 & 225,520 & 75,164 & 5,952 \\
  \rowcolor{cream}
  \cellcolor{almond} & Unique & 67,083 & 57,782 & 9,301 & 3,583 & 2,131 & 31,239 & 2,131 & 530 \\
  \rowcolor{almond}
  \multirow{-3}{*}{\textbf{Popularity}} & Avg & 31.35 & 19.09 & 12.27 & 7.22 & 6.36 & 94.80 & 31.59 & 2.50 \\

  \midrule
  \rowcolor{cream}
  & Aggregate & 93,769 & 55,154 & 38,615 & 20,612 & 18,571 & 187,827 & 54,918 & 5,533 \\
  \rowcolor{almond}
  \cellcolor{cream} & Unique & 67,195 & 55,115 & 12,080 & 3,108 & 1,769 & 28,935 & 1,769 & 530 \\
  \rowcolor{cream}
  \multirow{-3}{*}{\textbf{Popularity-Reverse}} & Avg & 30.64 & 18.02 & 12.62 & 6.74 & 6.07 & 77.49 & 22.66 & 2.28 \\

  \midrule
  \rowcolor{almond}
  \cellcolor{almond} & Aggregate & 84,266 & 49,315 & 34,951 & 17,610 & 16,041 & 214,779 & 59,605 & 6,591 \\
  \rowcolor{cream}
  \cellcolor{almond} &  Unique & 61,349 & 49,217 & 12,132 & 2,667 & 1,725 & 28,618 & 1,725 & 530 \\
  \rowcolor{almond} \multirow{-3}{*}{\textbf{Random}} & Avg & 28.73 & 16.81 & 11.92 & 6.00 & 5.47 & 83.31 & 23.12 & 2.56 \\

  \midrule
  \rowcolor{cream}
  & Aggregate & 83,718 & 51,180 & 32,538 & 17,338 & 15,455 & 201,862 & 68,040 & 6,820 \\
  \rowcolor{almond}
  \cellcolor{cream} & Unique & 61,421 & 51,150 & 10,271 & 2,823 & 1,665 & 27,876 & 1,665 & 536 \\
  \rowcolor{cream}
  \multirow{-3}{*}{\textbf{Random-Reverse}} & Avg & 28.07 & 17.16 & 10.91 & 5.81 & 5.18 & 79.38 & 26.76 & 2.68 \\

  \noalign{\vskip 1.1pt}
  \bottomrule
  \end{tabular}
  \centering
  \captionsetup{justification=centering}
  \caption{Cookie distribution across different measurements.}
  \label{table:cookies}

  \vspace{-5mm}

\end{table*}

\section{Cookie Expiration and Renewal}
\label{app:cookie-exp-dup}

To further explore the expiration time and the number of times cookies are set, we analyze cookies categorized as third-party, tracking, and \intractable for all four main runs. For the analysis, we use the unique cookies in a \cookiejar (\ie 'Unique' rows of \Cref{table:cookies}).

\vspace{1mm}

\begin{figure*}[]
     \centering
     \begin{subfigure}[b]{0.42\linewidth}
         \centering
         \includegraphics[width=\textwidth]{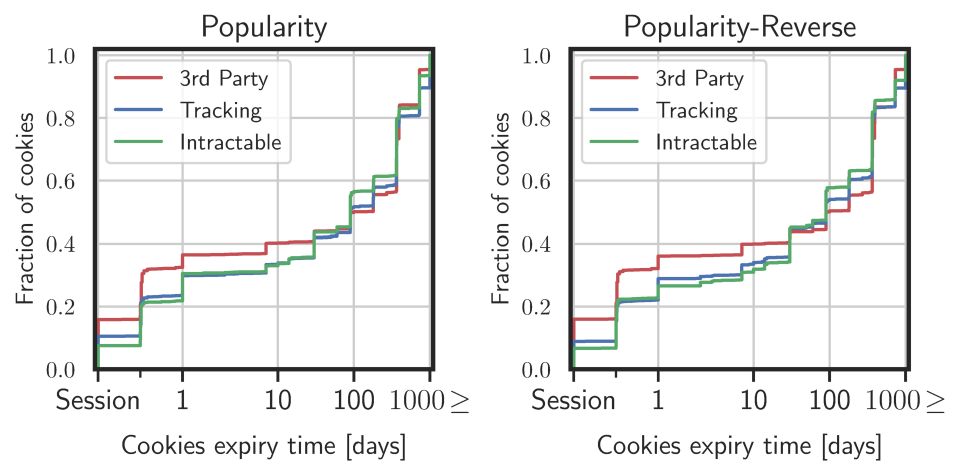}
         \caption{Popularity and Popularity-Reverse runs}
         \label{fig:cookie_expriry_regular}
     \end{subfigure}
          \begin{subfigure}[b]{0.42\linewidth}
         \centering
         \includegraphics[width=\textwidth]{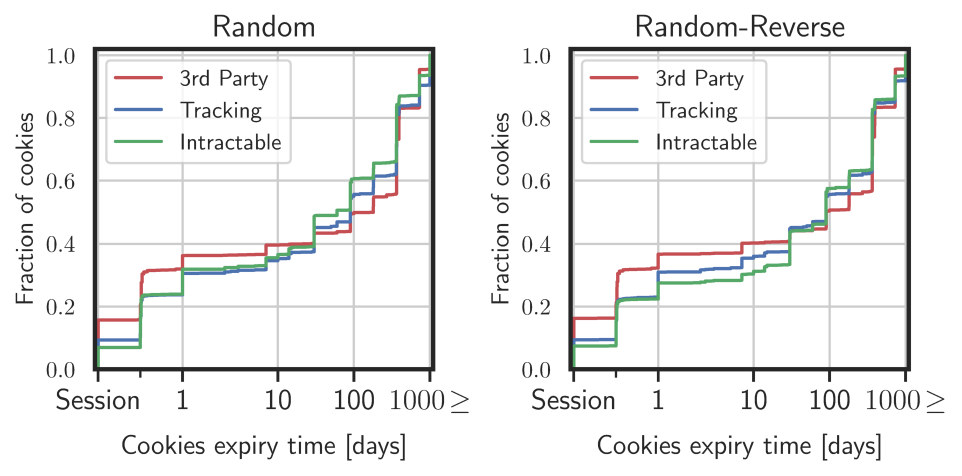}
         \caption{Random and Random-Reverse runs}
         \label{fig:cookie_expriry_random}
     \end{subfigure}
        \caption{Cookie Jar expiration time analysis.}
        \label{fig:ecdf_cookie_expiration}
\end{figure*}

\begin{figure*}[]
     \centering
     \begin{subfigure}{0.41\linewidth}
         \centering
         \includegraphics[width=\textwidth]{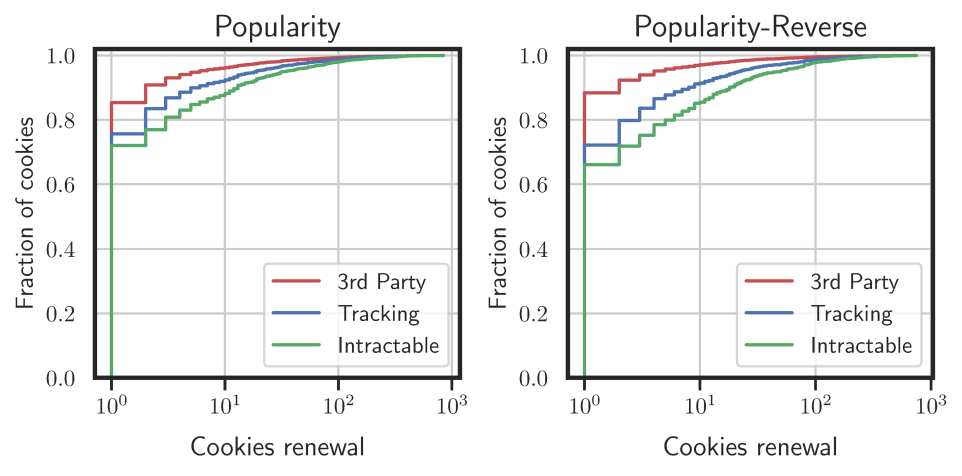}
         \caption{Popularity and Popularity-Reverse runs}
         \label{fig:cookie-duplication-regular}
     \end{subfigure}
          \begin{subfigure}{0.41\linewidth}
         \centering
         \includegraphics[width=\textwidth]{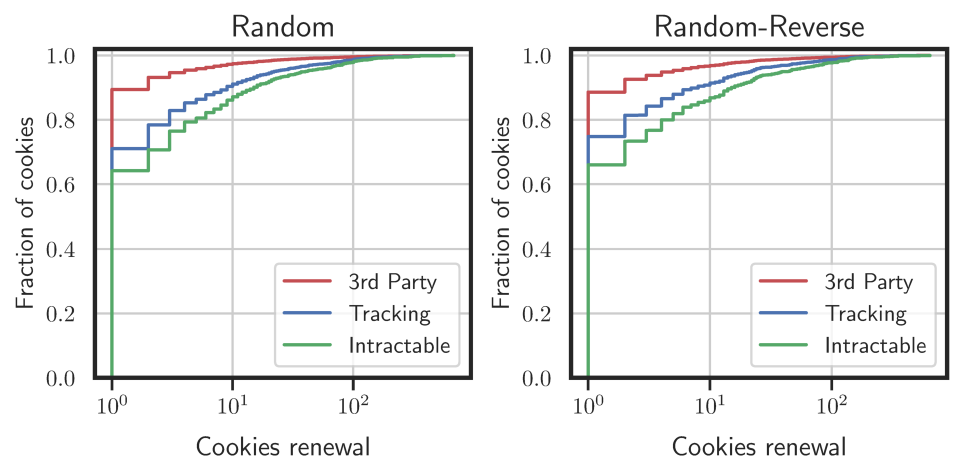}
         \caption{Random and Random-Reverse runs}
         \label{fig:cookie-duplication-random}
     \end{subfigure}
        \caption{Cookie Jar renewal analysis.}
        \label{fig:ecdf_cookie_duplication}
\end{figure*}

\parax{Expiration:} The ECDF graph displayed in Figure \ref{fig:ecdf_cookie_expiration} illustrates the distribution of cookies according to their expiry time in days. The label `Session' on the x-axis denotes cookies that are set to expire at the end of the browsing session. Across all categories and runs, we observe a consistent trend where nearly 60\% of cookies have an expiry exceeding 10 days. Furthermore, we note that the most common expiry time among these cookies is 365 days.
\vspace{1mm}

\parax{Renewal:} The ECDF graphs presented in \Cref{fig:ecdf_cookie_duplication} illustrate the distribution of cookies based on the number of times they are set or reset across different websites. We observe that \intractable cookies are more prone to being reset compared to tracking cookies, and even more so than third-party cookies, for all runs.
As shown in \Cref{fig:cookie_expriry_regular}, there is a subtle difference in the number of renewals of cookies across popularity runs. Specifically, the gap between graphs for different categories is larger in the \regularreverse run compared to \regular run. For instance, $\approx30\%$ of \intractable cookies are set more than once in \regular run, whereas this number is about 35\% for the \regularreverse run.  Nevertheless, the trends show similar patterns across all runs.

\end{document}